\documentclass[a4paper,11pt]{article}
 \pdfoutput=1
\usepackage[pdftex]{graphicx}
\usepackage{graphicx}
\usepackage[english]{babel}

\usepackage{amsmath}
\usepackage{amssymb}
\usepackage{color}
\usepackage{array} 
\newcolumntype{L}{>{\displaystyle}l} 
\newcolumntype{C}{>{\displaystyle}c} 
\newcolumntype{R}{>{\displaystyle}r} 
\usepackage{parskip}

\DeclareFontFamily{U}{mathx}{\hyphenchar\font45}
\DeclareFontShape{U}{mathx}{m}{n}{<-> mathx10}{}
\DeclareSymbolFont{mathx}{U}{mathx}{m}{n}
\DeclareMathAccent{\widebar}{0}{mathx}{"73}
\def\fatA{\mathbf{A}}
\def\fatv{\mathbf{v}}
\graphicspath{{../Figures/}{./Figures/}{../}{./}}
\makeatletter
\def\input@path{{./}{../}}
\makeatother

\def\bv{\bar{v}}
\def\fatdelta{\boldsymbol{\delta}}

\def\fatg{\mathbf g}
\def\fatn{\mathbf n}
\def\fatv{\mathbf v}
\def\fatI{\mathbf I}
\def\bh{\bar{h}}
\def\bH{\bar{H}}
\def\calD{\mathcal{D}}
\newcommand{\veps}{\varepsilon} 

\usepackage{float}
\usepackage{tikz}
\usepackage{color}
\usepackage{authblk}

\definecolor{hellgelb}{rgb}{1,1,0.85}
\definecolor{colKeys}{rgb}{0,0,1}
\definecolor{colIdentifier}{rgb}{0,0,0}
\definecolor{colComments}{rgb}{0.2,0.7,0}
\definecolor{colString}{rgb}{0.788,0,0.961}

\newcommand{\pder}[2]{\frac{\partial#1}{\partial#2}}

\newcommand{\ordo}[1]{\mathcal{O}(#1)}

\title{Accurate and stable time stepping in ice sheet modeling}
\author[1]{Gong Cheng}
\author[1]{Per L\"otstedt}
\author[1]{Lina von Sydow}

\affil[1]{Division of Scientific Computing, Department of Information Technology, Uppsala University, Uppsala, Sweden}

\date{}

\begin{document}

\maketitle

\begin{abstract}
In this paper we introduce adaptive time step control for simulation of evolution of ice sheets. The discretization error in the approximations is estimated using ``Milne's device'' by comparing the result from two different methods in a predictor-corrector pair. Using a predictor-corrector pair the 
expensive part of the procedure, the solution of the velocity and pressure equations, is performed only once per time step and an estimate of the local error is easily obtained. 
The stability of the numerical solution is maintained and the accuracy is controlled by keeping the local error below a given threshold using PI-control. Depending on the threshold, the time step 
$\Delta t$ is bound by stability requirements or accuracy requirements. 
Our method takes a shorter $\Delta t$ than an implicit method but with less work in each time step and the solver is simpler.
The method is analyzed theoretically with respect to stability and applied to the simulation of a 2D ice slab and a 3D circular ice sheet.
The stability bounds in the experiments are explained by and agree well with the theoretical results.  
\end{abstract}


\section{Introduction}
There is a growing interest in the prediction of the evolution of the large ice sheets on Antarctica and Greenland and their contribution to the future sea level rise
\cite{Alley, DeContoPollard, Hanna13, Vaughan}. Simulations of the dynamics of ice sheets in the past and in the future have been made, see e.g. \cite{PollardDeConto, Hakime2012}, 
but improvements in the modeling and the numerical methods are required for better fidelity, accuracy, and efficiency 
\cite{Stokesetal}. In this paper, we introduce a method to automatically
choose the time steps to control the discretization error and stability of the time integration of the governing system of partial differential equations (PDEs).

The full Stokes (FS) equations for the velocity field in the ice and an advection equation for the evolution of the ice surface 
are regarded as an accurate model of the motion of glaciers and ice sheets \cite{BlaHaftet, Hutter83, WGH99}. 
The viscosity in the FS equations depends non-linearly on the velocity. The numerical solution of the equations is
therefore demanding in terms of computational time. Hence, different simplifications of the FS equations have been derived under
various assumptions to reduce the computing effort. 
The shallow ice approximation (SIA) is based on the assumption that the thickness of the ice in the vertical direction is small compared to a length scale 
in the horizontal direction \cite{BlaHaftet}. Other approximations are the shallow shelf approximation (SSA) \cite{SSA, WGH99} 
and the Blatter-Pattyn model \cite{Blatter95, Pattyn2003}. Comparisons between solutions of the FS equations and the SIA equations are found in 
\cite{Ahlkrona13, Hindmarsh, ElmerVSsia}. The Ice Sheet Coupled Approximation Levels (ISCAL) is an adaptive method using SIA or FS in different parts of the ice sheet \cite{Ahlkrona16, Kirchner2016103}.

Numerical models have been implemented in codes for simulation of large ice sheets. They are often using a finite element method for the FS equations or 
approximations of them as in \cite{ElmerDescrip, Fabien2012, ISSM, LJGP, TPSTP} or a finite volume method as in \cite{Cornford13}. 
The PDE to evolve the thickness of the ice is time dependent and in the discretization of the time derivative a time step $\Delta t$ has to be chosen
for accuracy and stability.  
The stability of a class of one-step schemes with a $\theta$-parameter for the time derivative has been analyzed in \cite{Hind01}. Restrictions on $\Delta t$ are derived
by Fourier analysis of the linearized equations. If $\Delta x$ is the distance between the nodes in the space discretization then $\Delta t\le C_\ast\Delta x^2$ for some constant $C_\ast$.
These one-step schemes are applied to large ice sheets in \cite{IceThickness}.

The discretization of the PDE in space gives a system of ordinary differential equations (ODEs).
In the numerical solution of initial value problems for ODEs, the time step is often chosen to control the
estimated local error in the time discretization \cite{HNW, HW, Soder06}. Given the error estimate and the present time step, a new time step is selected
to the next time point such that an error tolerance is satisfied and the solution remains stable \cite{SoWa06}.

We introduce adaptive time step control for simulation of the ice sheet equations in the community ice sheet model Elmer/Ice \cite{ElmerDescrip}. 
Then the time step varies in the time interval of interest and there is no need to guess a stable and sufficiently accurate $\Delta t$ for the whole interval
in the beginning of the simulation. Spatial derivatives
are approximated by the finite element method in Elmer/Ice. The mesh is extruded in the vertical direction from a triangular or quadrilateral mesh in the horizontal plane.
It is adjusted in every time step to follow the free boundary at the ice surface. The dominant part of the computational effort is spent on the
solution of the equations for the velocity and the pressure in the ice.

The discretization error in the approximations is estimated using ``Milne's device'' by comparing the result from two different methods in a
predictor-corrector pair of Adams type of first and second order accuracy in time  \cite{Fujii, HNW}. The advantage with a predictor-corrector pair is that the 
expensive part of the procedure, the solution of the velocity and pressure equations, is performed only once per time step and that an estimate of the local error is
easily obtained. The time step $\Delta t$ is chosen to fulfill an error tolerance using PI control according to S{\"o}derlind \cite{SoWa06}. There is a bound on
$\Delta t$ depending on $\Delta x^2$ as in \cite{Hind01}. An unconditionally stable method would allow longer $\Delta t$ but also
require a fully implicit method and the solution of several different velocity equations in the iterations to compute the solution in every time step.

The outline of the paper is as follows. The equations that govern the evolution of the ice sheets are stated in Section \ref{sec:eq}. The predictor method is the Forward Euler method or the
second order Adams-Bashforth method and the corrector method is the Backward Euler method or the second order Adams-Moulton method (also referred to as the trapezoidal method) \cite{HNW}
or simplifications of them. The methods
are combined in Section \ref{sec:time} to solve for the velocities using FS, SIA, or ISCAL and the advection equation for the thickness. In Section \ref{sec:adap}, the time step control is introduced.
The stability of the methods applied to the thickness equation with the velocity from the SIA equation is analyzed as in \cite{Hind01} in Section \ref{sec:analysis}.
In Section \ref{sec:stab}, the stability of the predictor-corrector scheme is investigated.  
The time step control is tested in Section \ref{sec:numres} by simulation over long time intervals of examples in two and three dimensions from \cite{Ahlkrona16, Huybrechts, ISMIPHOM} using the
SIA, FS, and ISCAL solvers in Elmer/Ice \cite{Ahlkrona16, ElmerDescrip}. Conclusions are drawn in the final Section \ref{sec:conclusions}.

\section{Equations governing the ice sheet dynamics}
\label{sec:eq}
In this section we describe the equations and solvers for the flow of ice sheets.

\subsection{The full Stokes (FS) equations}
\label{sec:Stokes}
The flow of an ice sheet can be modeled by the non-linear FS equations \cite{Hutter83}. These equations are defined by balance of mass
\begin{equation}
\nabla \cdot{\bf{v}}=0,
\label{eq:mass}
\end{equation}
balance of momentum
\begin{equation}
\rho {\dot{{\fatv}}}=-\nabla p+ \nabla\cdot {\bf{T}}^D+\rho \fatg,
\label{eq:momentum}
\end{equation}
and a consitutive equation, the so called Glen's flow law 
\begin{equation}
{\bf{D}}=\mathcal{A}(T^{\prime})f(\sigma){\bf{T}}^D.
\label{eq:glen}
\end{equation}
Here $\bf{v}$ is the vector of velocities $\bf{v}=\left(\begin{array}{ccc} v_x&v_y&v_z\end{array}\right)^T$, $\rho$ is the density of the ice and $p$ is the pressure. The deviatoric stress tensor ${\bf{T}}^D$ is given by
\begin{equation}
{\bf{T}}^D=\left(\begin{array}{ccc}
t_{xx}^D&t_{xy}^D&t_{xz}^D\\
t_{yx}^D&t_{yy}^D&t_{yz}^D\\
t_{zx}^D&t_{zy}^D&t_{zz}^D\end{array}\right),
\label{eq:stress}
\end{equation}
where $t_{xx}^D$, $t_{yy}^D$, $t_{zz}^D$ and $t_{xy}^D$ denote longitudinal stresses and $t_{xz}^D$, $t_{yz}^D$ vertical shear stresses. We also have symmetry $t_{xy}=t_{yx}$, $t_{xz}=t_{zx}$ and $t_{yz}=t_{zy}$. The gravitational acceleration in the $z$-direction is denoted by $\bf{g}$, and the total time derivative of the velocity by ${\dot{\bf{v}}}$ which is very small and neglected in glaciological applications. Glen's flow law (\ref{eq:glen}) relates the stress and strain rate, where $\bf{D}=\frac{1}{2}\left(\nabla\bf{v}+(\nabla \bf{v})^T\right)$ is the strain rate tensor and $\mathcal{A}(T^\prime)$ the rate factor that describes how the viscosity depends on the pressure melting point corrected temperature $T^\prime$. For isothermal flow, the rate factor $\mathcal{A}$ is constant. Finally, 
\begin{equation}
f(\sigma)=\sigma^2
\label{eq:creep}
\end{equation}
is the creep response function for ice where $\sigma$ is the effective stress defined by 
\begin{equation}
\sigma^2=(t_{xz}^D)^2+(t_{yz}^D)^2+(t_{xz}^D)^2+\frac{1}{2}\left((t_{xx}^D)^2+(t_{yy}^D)^2+(t_{zz}^D)^2\right).
\label{eq:effectivestress}
\end{equation}

With the viscosity $\eta$ defined by
\begin{equation}
\eta=\left(2\mathcal{A}f(\sigma)\right)^{-1},
\label{eq:eta}
\end{equation}
the FS equations (\ref{eq:mass}), (\ref{eq:momentum}) and (\ref{eq:glen}) describing the flow of a non-Newtonian fluid can be written
\begin{equation}
\begin{array}{rcl}
\nabla\cdot(\eta(\nabla\fatv+\nabla\fatv^T))-\nabla p+\rho\fatg&=&0,\\
\nabla\cdot\fatv&=&0.
\end{array}
\label{eq:FS}
\end{equation}

If the ice base is frozen, the velocity $\mathbf{v}$ satisfies a no slip condition at the base 
\begin{equation}
\fatv=0.
\label{eq:noslip}
\end{equation} 
An ice sliding at the base is modeled by a sliding law  \cite{Schoof}.
Let $\fatI$ be the identity matrix. At the surface with normal $\fatn$, the ice is stress free with
\begin{equation}
(-p\fatI+{\bf{T}}^D)\cdot\fatn=0.
\label{eq:stressfree}
\end{equation} 

The FS equations \eqref{eq:mass} and \eqref{eq:FS} are discretized in space by a finite element method with linear P1-P1 elements with stabilization 
to avoid spurious oscillations in the pressure using the standard setting in Elmer/Ice \cite{ElmerDescrip}. The resulting system of non-linear equations is solved by
Newton iterations. The system of linear equations in every Newton iteration is solved iteratively.

\subsection{The Shallow Ice Approximation (SIA)} 
\label{sec:SIA}
The SIA is derived from FS by scaling of variables and perturbation expansions, see e.g. \cite{BlaHaftet, GreveBlatterBok}. The underlying assumption is that the aspect ratio $\varepsilon$ of the ice sheet -- the quotient between the thickness $H$ and a length scale $L$ of the ice sheet -- is small. The SIA velocities and pressure can be computed from the following expressions (using $v_{xb}$ and $v_{yb}$ as the basal sliding velocities, the Euclidean vector norm $\|\cdot\|_2$, and $g=\|\fatg\|_2$)
\begin{equation}
\begin{array}{rcl}
v_x&=&{\displaystyle{v_{xb}-2(\rho g)^3\frac{\partial h}{\partial x}\|\nabla h\|_2^2\int_b^z\mathcal{A}(h-s)^3ds,}}\\
\\
v_y&=&{\displaystyle{v_{yb}-2(\rho g)^3\frac{\partial h}{\partial y}\|\nabla h\|_2^2\int_b^z\mathcal{A}(h-s)^3ds,}}\\
\\
v_z&=&{\displaystyle{-\int_b^z \left(\frac{\partial v_x}{\partial x}+\frac{\partial v_y}{\partial y}\right) ds,}}\\
\\
p&=&{\displaystyle{\rho g(h-z).}}\end{array}
\label{eq:SIA}
\end{equation}
In \cite{Ahlkrona13, Ahlkrona13a} the validity and accuracy of SIA were examined. Due to the assumptions in the derivation of SIA it does not perform well in regions with large spatial variations in data, at steep margins, in ice streams, in ice shelves, and at domes.

\subsection{The Ice Sheet Coupled Approximation Levels (ISCAL)}
\label{sec:ISCAL}
While FS is an accurate model for ice sheet flow, it is computationally expensive to solve. SIA on the other hand is computationally cheap, but fails to compute accurate solutions in large regions of the ice sheet for realistic glaciological applications. For this reason, FS and SIA were coupled into ISCAL and implemented in Elmer/Ice in \cite{Ahlkrona16}. This method decides automatically and dynamically where SIA is valid and use this approximation in these regions. FS is employed for the remaining part of the ice sheet where a more accurate solver is required. This way the overall computational complexity is substantially reduced compared to FS, still being much more accurate than SIA. ISCAL was applied to a simplified ice sheet covering Svalbard in \cite{Kirchner2016103}. 
\subsection{The free surface equation and the thickness equation}
\label{sec:height}
\begin{figure}[H]
\center
\includegraphics[width=12cm]{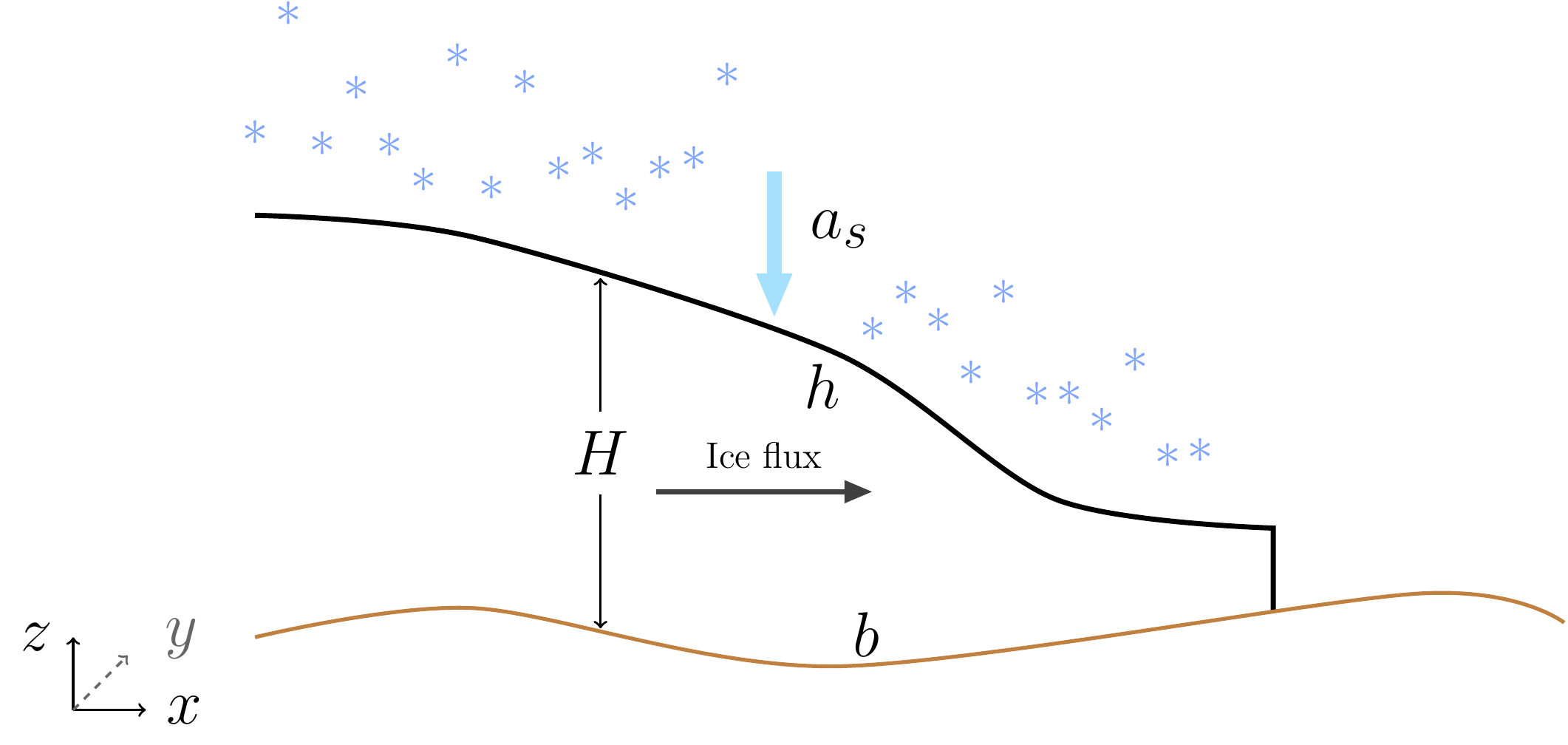} 
\caption{An ice sheet with height $h$, bedrock elevation $b$, thickness $H$, and accumulation $a_s$.}
\label{fig:ice}
\end{figure}
The $z$-coordinate of the free surface position $h(x,y,t)$ (see Figure \ref{fig:ice}) is given by the free surface equation
\begin{equation}
  \pder{h}{t}+v_x\pder{h}{x}+v_y\pder{h}{y}-v_z=a_s,
\label{eq: free surface}
\end{equation}
where $a_s$ denotes the net surface accumulation/ablation. As an alternative to solving this equation for $h(x,y,t)$, we can solve the thickness advection equation 
\begin{equation}
  \pder{H}{t}+\pder{q_x}{x}+\pder{q_y}{y}=a_s
  \label{eq: Thickness}
\end{equation}
for $H(x,y,t)=h(x,y,t)-b(x,y)$ (see Figure \ref{fig:ice}). The $z$-coordinate of the ice base is $b(x, y)$. In \eqref{eq: Thickness}, the flux is
\begin{equation}
q_x=\int_b^h v_x\,dz,\quad q_y=\int_b^h v_y\,dz.
\label{eq:q}
\end{equation}
Both the free surface \eqref{eq: free surface} and the thickness \eqref{eq: Thickness} equation are solved in one dimension lower than the velocity equation. 
In this paper, we will use the thickness equation (\ref{eq: Thickness}) to compute the time evolution of the ice sheet. 
A stabilization term is added to the equation making the spatial discretization behave like an upwind scheme according to the direction of the velocity.

\section{Time stepping}
\label{sec:time}
We will use a predictor-corrector time stepping scheme for (\ref{eq: Thickness}), rewritten as
\begin{equation}
  \pder{H}{t}=a_s-\pder{q_x}{x}-\pder{q_y}{y}=f(H,v).\label{eq:Thickness2}
\end{equation}
The numerical approximation of $H$ at time $t_n, \, n\ge 0,$ is $H^n$ and the time step is $\Delta t_n=t_n-t_{n-1}$. 

The predictor-corrector algorithm becomes
\begin{enumerate}  
    \item {\bf Predictor step}: Solve  for $\tilde{H}^{n}$ explicitly in time from $v^{n-1}$, $H^{n-1}$, $H^{n-2}$, etc.
    \item {\bf Velocity solver}: Solve for $v^n$ using the predictor  $\tilde{H}^{n}$.
    \item {\bf Corrector step}: 
    \begin{itemize}
	    \item Fully implicit scheme: Solve for $H^n$ implicitly from  $v^n$, ${H}^{n}$, $H^{n-1}$, $H^{n-2}$, etc.
	    \item Semi-implicit scheme: Solve for $H^n$ implicitly from  $v^n$, $\tilde{H}^{n}$, $H^{n-1}$, $H^{n-2}$, etc.
	\end{itemize}

\end{enumerate} 
The velocities in Step 2 can be computed using either FS, SIA, or ISCAL. 
  
  We consider both a first order predictor step using Forward Euler (FE)
    \begin{equation}
        \tilde{H}^{n}=H^{n-1}+\Delta t_n f(H^{n-1},v^{n-1}),
        \label{eq:FE}
        \end{equation}
and the second order Adams-Bashforth method (AB)
    \begin{equation}
      { \tilde{H}^{n}}={H^{n-1}}+\Delta t_n\left[\beta^n_1f({H^{n-1}},v^{n-1})+\beta^n_2f({H^{n-2}},v^{n-2})\right],
    \label{eq:AB}
    \end{equation}   
    where 
    \[
      \zeta^n=\frac{\Delta t_n}{\Delta t_{n-1}},\quad\beta^n_1=1+\frac{\zeta^n}{2},\quad\beta^n_2=-\frac{\zeta^n}{2}.
    \]
For the corrector step, the fully implicit and semi-implicit Backward Euler are considered as first order accurate methods
\begin{equation}
        {H}^{n}=H^{n-1}+\Delta t_n f({{H}^{n}},v^{n}),
               \label{eq:FBE}
 \end{equation}
        
        \begin{equation}
        {H}^{n}=H^{n-1}+\Delta t_n f({\tilde{H}^{n}},v^{n}),
        \label{eq:SBE}
        \end{equation}
        denoted (FBE) and (SBE), respectively. The fully implicit Adams-Moulton (FAM) 
             \begin{equation}
      {H^{n}}={H^{n-1}}+\frac{\Delta t_n}{2}\left[ f({ {H}^{n}},v^{n}) + f({H^{n-1}},v^{n-1})\right],
             \label{eq:FAM}
    \end{equation}
        and semi-implicit Adams-Moulton (SAM)         
            \begin{equation}
      {H^{n}}={H^{n-1}}+\frac{\Delta t_n}{2}\left[ f({ \tilde{H}^{n}},v^{n}) + f({H^{n-1}},v^{n-1})\right],
    \label{eq:SAM}
    \end{equation} 
    are the second order methods considered. 
    
    Four different predictor-corrector schemes are listed in Table \ref{tab:schemes}.
    \begin{table}[H]
    \begin{centering}
    \begin{tabular}{|l|l|l|}
    \hline
    Scheme&Predictor&Corrector\\
    \hline
    FE-FBE&FE, Equation (\ref{eq:FE})&FBE, Equation (\ref{eq:FBE})  \\
    FE-SBE&FE, Equation (\ref{eq:FE})&SBE, Equation (\ref{eq:SBE})  \\
    AB-FAM&AB, Equation (\ref{eq:AB})&FAM, Equation (\ref{eq:FAM})  \\ 
    AB-SAM&AB, Equation (\ref{eq:AB})&SAM, Equation (\ref{eq:SAM}) \\ \hline 
    \end{tabular}
    \caption{The four predictor-corrector schemes considered.}    \label{tab:schemes}

    \end{centering}
    \end{table}

The first time step at $t=0$ is taken by the first order method. In all the other time steps, the solution is advanced by the first or the second order method.
    
    The schemes FE-FBE and AB-FAM are only used in the analysis, see Section \ref{sec:stab}. In Section \ref{sec:numres}, numerical results 
using FE-SBE and AB-SAM are presented.

\section{Time step control}
\label{sec:adap}
In each time step, we will compute a new $\Delta t_{n+1}$ with the following algorithm:
\begin{itemize}
    \item {\bf Error estimate}: Estimate the local trunction error $\tau^n$.
    \item {\bf Adaptive time step}: Compute $\Delta t_{n+1}$ from $\Delta t_n$, the local truncation errors $\tau^n$, $\tau^{n-1}$, and a given tolerance $\epsilon$ using a PI controller from \cite{SoWa06}.
    \end{itemize}

	FE (\ref{eq:FE}) has the local error between the analytical solution $H(t_n)$ with initial data $H^{n-1}$ and the numerical solution 
	\begin{equation}
		\begin{array}{rcl}
		H(t_n)-\tilde{H}^{n}=c_P\Delta t^2_nH''(t_n)+\mathcal{O}(\Delta t_n^3)&,&
		c_P=\frac{1}{2},
		\end{array}
		\label{eq:errFE}
	\end{equation}  
	while SBE (\ref{eq:SBE}) has the local error
	\begin{equation}
      \begin{array}{rcl}
        H(t_n)-H^{n}=c_I \Delta t^2_nH''(t_n)+\mathcal{O}(\Delta t_n^3)&,&
        c_I =-\frac{1}{2}.
      \end{array}
        \label{eq:errSBE}
    \end{equation}   
Combining (\ref{eq:errFE}) and (\ref{eq:errSBE}) gives the following leading term of the local truncation error for SBE
    \begin{equation}
      \tau^{n}=\frac{c_I(H^{n}-\tilde{H}^{n})}{\Delta t_n(c_I-c_P)}=\frac{1}{2\Delta t_n}(H^{n}-\tilde{H}^{n}).
      \label{eq:tau1}
    \end{equation}
From (\ref{eq:tau1}) we compute the next time step using PI.4.2 in \cite{SoWa06}
    \begin{equation}
      \Delta t_{n+1}=\left(\frac{\epsilon}{\eta_n}\right)^{\beta_1}\left(\frac{\epsilon}{\eta_{n-1}}\right)^{\beta_2}\Delta t_{n},\; n=1,2,\ldots,
     \label{eq: dt2}
   \end{equation}

    where ${\displaystyle{\eta=\max_{\Omega}{|\tau^{n}|}}}$ over the spatial domain $\Omega$ with parameters $\beta_1=3/10$ and $\beta_2=-1/10$ suggested in \cite{SoWa06}.

Similarily, we have for the second order method that AB in (\ref{eq:AB}) has local error
    \begin{equation}
      \begin{array}{rcl}
        H(t_n)-{ \tilde{H}^{n}}=c_P(\zeta^n)\Delta t^3_nH'''+\mathcal{O}(\Delta t_n^4)&,
        &c_P(\zeta^n)=\frac{1}{6}+\frac{1}{4\zeta^n},
      \end{array}
    \end{equation} 
    and SAM in (\ref{eq:SAM}) has local error
    \begin{equation}
      \begin{array}{rcl}
        H(t_n)-{H^{n}}=c_I \Delta t^3_nH'''+\mathcal{O}(\Delta t_n^4)&,&
        c_I =-\frac{1}{12}.
      \end{array}
    \end{equation}
    Again, combining the expressions for the local errors gives the following local truncation error for SAM
    \begin{equation}
      \tau^n=\frac{c_I({H^{n}}-{ \tilde{H}^{n}})}{\Delta t_n(c_I-c_P)}=\frac{\zeta^n(H^{n}-\tilde{H}^{n})}{(3\zeta^n+3)\Delta t_n},
      \label{eq:tau2}
    \end{equation}
and the new time step using PI.4.2 is given by \eqref{eq: dt2} with $\beta_1=1/5$ and $\beta_2=-1/15$, see \cite{SoWa06}.

\section{Analysis of a simplified 2D-problem}
\label{sec:analysis}
A stability analysis for a 2D-problem is performed using SIA in this section. The analysis follows \cite{Hind01}, but our final results are slightly more comprehensive.

\subsection{Analytical expressions}
\label{sec:lin}

For a 2D-problem, we derive by (\ref{eq:SIA}) and the no-slip condition that the SIA-velocities are given by
\begin{equation} 
  \begin{split}
    v_x& = -\frac{1}{2}{\mathcal{A}}(\rho g)^3\left(\pder{h}{x}\right)^3\left(H^4-(h-z)^4\right),\\
    v_z& =\frac{1}{2}{\mathcal{A}}\left(\rho g\right)^3\left\{4\left(\pder{h}{x}\right)^3\left[H^3\pder{H}{x}(z-b)+\frac{1}{4}\pder{h}{x}\left((h-z)^4-H^4\right)\right]\right.+\\
      & \left.+3\left(\pder{h}{x}\right)^2 \left[H^4(z-b)-\frac{1}{5}\left(H^5-(h-z)^5\right) \right] \frac{\partial^2 h}{\partial x^2}\right\}.  
  \end{split}
  \label{eq:vSIA}
\end{equation}

The average velocity in the vertical direction is denoted by $\bar v$. Use \eqref{eq:q} and $C=\frac{2}{5}{\mathcal{A}}(\rho g)^3$ to obtain
\begin{equation} 
 \bar{v}=\frac{q_x}{H}=\frac{\int_b^hv_x\,dz}{H}=-CH^4\left(\pder{h}{x}\right)^3. \label{eq: average velocity}
\end{equation}

Using this in (\ref{eq: Thickness}) gives
\begin{equation} 
 \pder{H}{t}+\pder{(\bar{v}H)}{x}=a_s,\label{eq: thickness solved in Elmer}
\end{equation}
which can also be written as an equation related to a $p$-parabolic equation with $p=4$ \cite{DiBenedetto}
\begin{equation}\label{eq: h_compact}
  \pder{H}{t}-\pder{}{x}\left(CH^5\left|\pder{h}{x}\right|^2\pder{h}{x}\right)=a_s.
\end{equation}

In general, an ice sheet model is a coupled system consisting of equations for velocities, thickness, temperature, grounding-line migration, and bedrock motion. Numerically, 
these equations are usually solved separately in a time step keeping the variables from the other equations constant. For instance, we use a fixed $H$ at the current time step when 
we solve for the velocity $\bar{v}$ and then use the fixed $\bar{v}$ to solve for the evolution of the thickness $H$. In the analysis of the time discretization of the
thickness equation \eqref{eq: h_compact}, there are two different sources of $H$: one from the calculation of the velocity $\bar{v}$ denoted by $\hat{H}$ and one from the integration of the 
thickness equation written $H$. Then, the new thickness equation is
\begin{equation}\label{eq: H seperated}
  \pder{H}{t}-\pder{}{x}\left(C{\hat{H}}^4\left|\pder{\hat{h}}{x}\right|^2\pder{\hat{h}}{x} H\right)=a_s.
\end{equation}

The coupled system \eqref{eq: H seperated} is linearized by introducing the perturbation $\delta(x,t)$ about the steady state solution for the 
thickness $\bH(x)$ and the height $\bar{h}(x)$. 
Then $H$ and $\hat{H}$ are expressed as
\begin{equation}
  \begin{split}
  H = \bH+\delta=\bh-b+\delta,\\
  \hat{H} = \bH+ \hat\delta=\bh-b+\hat\delta,
  \end{split}
\label{eq:Hperturbed}
\end{equation}
and after ignoring quadratic terms and higher in $\delta$ we arrive at
\begin{equation}
  \begin{split}
    \pder{\delta}{t}&=C\pder{}{x}\left[4\bH^4\left(\pder{\bh}{x}\right)^3\hat{\delta}+3 \bH^5\left(\pder{\bh}{x}\right)^2\pder{\hat{\delta}}{x}+\bH^4\left(\pder{\bh}{x}\right)^3\delta\right],
  \end{split}\label{eq: perturbation eq}
\end{equation}
where the first two terms on the right hand side derive from $\bar{v}$, cf. \cite{Hind01}.

\begin{figure}[H]
\center
\includegraphics[width=6cm]{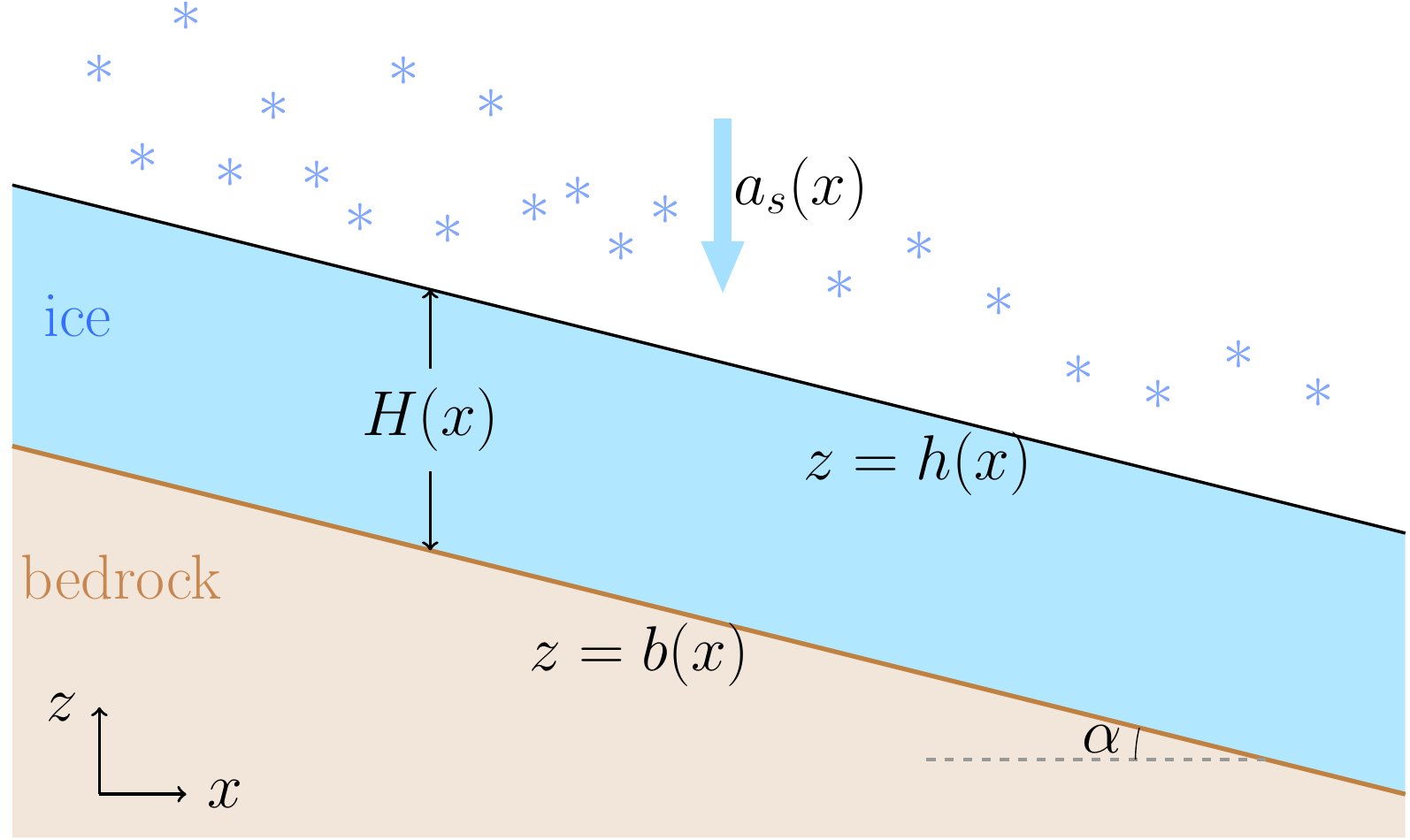} 
\caption{A slab-on-slope with time independent $H$ and $h$.}
\label{fig:slab}
\end{figure}
In a slab-on-slope case (see Figure \ref{fig:slab}), we have a constant $\bH$ and $\pder{b}{x}=\pder{\bh}{x}=\alpha$ and the model equation for $\delta$ is
\begin{equation} 
  \pder{\delta}{t}=4C\alpha^3\bH^4\pder{\hat{\delta}}{x}+3C\alpha^2\bH^5\pder{}{x}{\left(\pder{\hat{\delta}}{x}\right)}+C\alpha^3\bH^4\pder{\delta}{x},
  \label{eq:final thickness}
\end{equation}
Since the coefficient in front of the second derivative is positive, the solution $\delta$ is stable.
In the next section, the thickness equation is discretized by an upwinding scheme and central differences are used for the derivative in the velocity solution.
This is an equation modeling the time dependence of $H$ in Elmer/Ice. The same result is obtained if \eqref{eq: H seperated} is first discretized and then linearized.

\subsection{Stability analysis}
\label{sec: stab analysis}
The stability is investigated here when (\ref{eq:final thickness}) is discretized in time using a $\theta$-method with the time step $\Delta t$. When $\theta=0$, the $\theta$-method is the Forward Euler method and for $\theta>0$, it is a mixed implicit/explicit method. In this analysis, $\hat{\delta}$ is always evaluated at the current time point.

First, we study Forward Euler in time with $\alpha<0$ and centered and upwinding differences in space with step size $\Delta x$
as described in Section \ref{sec:lin}. Let $\delta_j^n$ approximate $\delta(x_j, t_n)$ where $x_j=x_{j-1}+\Delta x$ and $t_{n}=t_{n-1}+\Delta t$. Then
\begin{equation} 
  \frac{\delta_j^{n+1}-\delta_j^{n}}{\Delta t}=5C\alpha^3\bH^4\frac{\delta_j^{n}-\delta_{j-1}^{n}}{\Delta x}+3C\alpha^2\bH^5\frac{\delta_{j+1}^{n}-\delta_{j-1}^{n}-\delta_{j}^{n}+\delta_{j-2}^{n}}{2\Delta x^2}.
\end{equation}

Introducing  
\begin{equation}
  \begin{array}{rcccl}
    \xi=\frac{\Delta t}{\Delta x}&,&a=5C\alpha^3\bH^4\xi&,&
    c=\frac{3}{2}C\alpha^2\bH^5\frac{\Delta t}{\Delta x^2},
  \end{array}\label{eq: coeff a and c}
\end{equation}
we arrive at
\begin{equation}
\delta_j^{n+1}=\delta_j^{n}+a\left(\delta_j^{n}-\delta_{j-1}^{n}\right)+c\left(\delta_{j+1}^{n}-\delta_{j}^{n}-\delta_{j-1}^{n}+\delta_{j-2}^{n}\right).
\label{eq:G}
\end{equation}
For the case $\alpha>0$,
\begin{equation}
\delta_j^{n+1}=\delta_j^{n}+a\left(\delta_{j+1}^{n}-\delta_{j}^{n}\right)+c\left(\delta_{j+2}^{n}-\delta_{j+1}^{n}-\delta_{j}^{n}+\delta_{j-1}^{n}\right).
\label{eq:G2}
\end{equation}

Replacing $\delta_k^n$ by the Fourier modes ${\delta}_\omega^ne^{ik\omega \Delta x}$ gives ${\delta}_{\omega}^{n+1}=\lambda {\delta}_{\omega}^{n}$ where $\lambda$ is
\begin{equation} 
  \begin{array}{rcl}
  \lambda &= &1+|a|(\cos{\omega\Delta x}+i\sin{\omega\Delta x}-1)+\\
    &+&c(\cos{2\omega\Delta x}+i\sin{2\omega\Delta x}-2i\sin{\omega\Delta x}-1),
  \end{array}
\label{eq: growth factor 2}
\end{equation}
considering both cases $\alpha<0$ and $\alpha>0$.
For stability in the time discretization \eqref{eq:G} and \eqref{eq:G2} for a given $\Delta x$, the requirement on $\lambda$ is $|\lambda|\le 1$
for all $\omega\Delta x\in[0, \pi]$. Let
\begin{equation}  
  \displaystyle{k=\frac{|a|}{c}=\frac{10}{3}|\alpha|\frac{\Delta x}{\bH}}.
\label{eq:kdef}
\end{equation}
Then the restriction on the time step is (for details, see \ref{app:a})
\begin{equation}  
  \Delta t\leq\frac{k+2+2\sqrt{k}}{2k^2+8}\frac{2\Delta x^2}{3C\alpha^2\bH^5}.
\label{eq:dtrestr}
\end{equation}

SIA is accurate when a typical length scale $L$ in the horizontal direction is large compared to $\bH$ such that $\bH=\veps L$ with a small $\veps$ \cite{Ahlkrona13}.
Then $k=10|\alpha|\Delta x/(3\veps L)$. When $\veps\propto 0.01, \Delta x/L\propto 0.01-0.1$, and $\alpha\propto 0.01-0.1,$ $k$ will be $\propto 0.01-1$ 
and the factor with $k$ in \eqref{eq:dtrestr}
is $\propto 1$, the bound on $\Delta t$ will decrease with $\Delta x^2$ as expected with an explicit discretization of a parabolic equation \eqref{eq: h_compact}
and decreases rapidly with increasing thickness as $\bH^{-5}$. Only when $k$ is large in \eqref{eq:kdef}, e.g. when the ice is very thin, 
we have $\Delta t\le C_1 \Delta x/\bH^4$ and longer time steps are possible.  Compared to the bound in \cite{Hind01}, the bound in (\ref{eq:dtrestr}) is sharp and more detailed.

A blend of the spatial first derivatives at two time levels is defined by a $\theta$-parameter.
Using the same notation as in \eqref{eq: coeff a and c} for $a$ and $c$, 
the fully discretized scheme is for (\ref{eq:final thickness}) with $\alpha<0$
\begin{equation}
  \begin{split}
    \delta_{j}^{n+1}-\delta_{j}^{n}=&\ c\left(\delta_{j+1}^{n}-\delta_{j-1}^{n}-\delta_{j}^{n}+\delta_{j-2}^{n}\right)\\
    &+a\left[(1-{\theta})\left(\delta_j^n-\delta_{j-1}^n\right)+{\theta}\left(\delta_j^{n+1}-\delta_{j-1}^{n+1}\right)\right].
  \end{split}
\label{eq:secondorder}
\end{equation}
The range of $\theta$ is $[0,\frac{1}{5}]$ since the first derivatives of $\delta$ are from two difference sources. 
As shown in \eqref{eq:final thickness}, $\pder{\hat{\delta}}{x}$ in the first and second terms of the right hand side are computed explicitly in 
the velocity equation and discretized in the previous time step. Therefore, only $1/5$ of the first derivatives are determined by the $\theta$-method. 
The remaining $4/5$ of the first derivatives and the second derivative are always discretized explicitly in time. 
An implicit method, e.g. the Backward Euler method, applied only to the thickness equation is not a fully implicit method for the coupled system.


Again using Fourier analysis gives for the growth factor with $\theta\in[0,\frac{1}{5}]$
\begin{equation}\label{eq:lambda4}
  \begin{array}{rl}
    \lambda_\theta=&\left[1+c(\cos{2\omega\Delta x}+i\sin{2\omega\Delta x}-2i\sin{\omega\Delta x}-1)\right.\\
                 &\left.+|a|(1-{\theta})(\cos{\omega\Delta x}+i\sin{\omega\Delta x}-1)\right]\\
                 & /\left[1-|a|{\theta}(\cos{\omega\Delta x}+i\sin{\omega\Delta x}-1)\right].
  \end{array}
\end{equation}
Analytical bounds for $\Delta t$ in \eqref{eq:lambda4} cannot be derived as easily as for Forward Euler in \eqref{eq:dtrestr}  when $\theta=0$ in \eqref{eq:secondorder}. Using the notation $\lambda$ in \eqref{eq: growth factor 2}, the growth factor for $\theta$-method \eqref{eq:lambda4} is now
\begin{equation}\label{eq:lambda4 simplified}
  \begin{split}
    \lambda_\theta=\frac{\lambda-|a|{\theta}(\cos{\omega\Delta x}+i\sin{\omega\Delta x}-1)}{1-|a|{\theta}(\cos{\omega\Delta x}+i\sin{\omega\Delta x}-1)}.
  \end{split}
\end{equation}
Let $z=|a|{\theta}(\cos{\omega\Delta x}+i\sin{\omega\Delta x}-1)$ and $\bar{z}$ is the complex conjugate of $z$. By simple calculation with the assumption above, we know that $|z|\ll1$ and $\Re z\leq 0$. Then,
\begin{equation}\label{eq:lambda4 simplified with z}
    |\lambda_\theta|=\left|\frac{\lambda-z}{1-z}\right|=\frac{|\lambda+|z|^2-(z+\lambda\bar{z})|}{1-(z+\bar{z})+|z|^2}\leq|\lambda|+|z|^2+|z|(1+|\lambda|).
\end{equation}
The bound on a stable $\Delta t$ for the $\theta$-method is is in most cases more restricted than the bound for the explicit method. The exact bounds on $\Delta t$ for 
stability for the $\theta$-method can be computed numerically. 

The separated procedure for velocity and thickness is a typical way of solving the coupled system. However, this is potentially problematic in ice 
sheet modeling since it tends to generate a diffusion term which is always solved explicitly in time. In Section \ref{sec:numres1}, we compare numerical experiments with this analysis.

\section{Stability analysis for the predictor-corrector schemes}
\label{sec:stab}
The stability for (\ref{eq: Thickness}) is analyzed here using the full predictor-corrector time stepping scheme in Section \ref{sec:time}. 
The assumption is that $q_y=0$ in \eqref{eq:q} and that $q_x=\bv H$ as in \eqref{eq: average velocity}. 
The finite element discretization of the space derivative in \eqref{eq: Thickness} with linear hat functions in 1D is stabilized by adding a term proportional to the
first derivative squared.  On an equidistant mesh, the approximation corresponds to a finite difference discretization with upwinding for $\partial (\bv H)/\partial x$.

Consider the predictor-corrector scheme AB-FAM, i.e. $\tilde{H}^n$ is computed from (\ref{eq:AB}), $\bv^n=\bv^n(\tilde{H}^n)$, and 
$H^n$ is computed by (\ref{eq:FAM}) (see Table \ref{tab:schemes}). 
Assuming that $\bv>0$, $a_s$ is constant in time, and that $\bH^n$ is the steady state solution yields 
$\frac{\partial q^n_x}{\partial x}=\frac{\partial (\bar{v}^n\bH^n)}{\partial x}=a_s$ and consequently
\begin{equation}\label{eq:assump}
\bv_j^{k}\bH_j^k-\bv_{j-1}^{k}\bH_{j-1}^k=\Delta x a_s,\; k=n, n-1.
\end{equation}
Then by FAM $\bH_{j}^{n+1}=\bH_j^n$ and by AB $\tilde{H}_{j}^{n+1}=\bH_j^n=\bH_j^{n-1}=\tilde{H}_{j}^{n}$. Introduce as in Section \ref{sec: stab analysis} 
perturbations $\delta_j^n$ to the steady state at $x_j$ denoted by $\bH_j$ in order to analyze the stability for these perturbations. FAM for $\bH_j+\delta_j^{n+1}$ gives with $\xi$ defined in \eqref{eq: coeff a and c}
\begin{equation}\label{eq:BDF2lini}
\begin{array}{l}
 (\bH_j+\delta_j^{n+1})-(\bH_j+\delta_j^{n})+\\
  \frac{1}{2}\xi\left(\bv_j(\bH+\tilde{\delta}^{n+1})(\bH_j+\delta_j^{n+1})-\bv_{j-1}(\bH+
      \tilde{\delta}^{n+1})(\bH_{j-1}+\delta_{j-1}^{n+1})\right)+\\
   \frac{1}{2}\xi\left(\bv_{j}(\bH+\tilde{\delta}^{n})(\bH_j+\delta_j^{n})
   -\bv_{j-1}(\bH+\tilde{\delta}^{n})(\bH_{j-1}+\delta_{j-1}^{n})\right)=a_s.
\end{array}
\end{equation}

Linearize \eqref{eq:BDF2lini} ignoring quadratic terms in $\delta$. We will also assume as in SIA that
$\bv_{j}(H)$ depends only locally on $H_{j}$ as in \eqref{eq:SIA}, see \ref{app:b}.
Then use assumption \eqref{eq:assump} to arrive at
\begin{equation}\label{eq:BDF2lini2}
\begin{array}{l}
 \delta_j^{n+1}-\delta_j^{n}+\\
   \frac{1}{2}\xi\left(\bv(\bH_{j})\delta_j^{n+1}-\bv(\bH_{j-1})\delta_{j-1}^{n+1}
   +\frac{\partial \bv}{\partial H_j}(\bH_{j})\bH_j\tilde{\delta}_{j}^{n+1}
          -\frac{\partial \bv}{\partial H_{j-1}}(\bH_{j-1})\bH_{j-1}\tilde{\delta}_{j-1}^{n+1}\right)+\\
   \frac{1}{2}\xi\left(\bv(\bH_{j})\delta_j^{n}-\bv(\bH_{j-1})\delta_{j-1}^{n}
   +\frac{\partial \bv}{\partial H_j}(\bH_{j})\bH_j\tilde{\delta}_{ j}^{n}-\frac{\partial \bv}{\partial H_{j-1}}(\bH_{j-1})\bH_{j-1}\tilde{\delta}_{ j-1}^{n}\right)=0.
\end{array}
\end{equation}

Assume that $\bv$ and $\frac{\partial \bv}{\partial H_j}$ vary smoothly in space and time such that $\bv(\bH_j)-\bv(\bH_{j-1})$ and 
$\frac{\partial \bv}{\partial H_j}(\bH_j)-\frac{\partial \bv}{\partial H_{j-1}}(\bH_{j-1})$ are small. 
Introducing
\begin{equation}
\begin{array}{rcl}
\mu&=&\frac{\partial \bv}{\partial H_j}(\bH_j)(1-e^{-i\omega\Delta x}),\\
\nu&=&\bv(\bH_j)(1-e^{-i\omega\Delta x}),\end{array}
\label{eq:munu}
\end{equation}
where $\omega\Delta x\in[0, \pi]$, 
and replacing $\delta_j^n$ and $\tilde{\delta}_{ j}^n$ by the Fourier modes 
$\delta_{\omega}^ne^{ij\omega\Delta x}$ and $\tilde{\delta}_{\omega}^n e^{ij\omega\Delta x}$ in  \eqref{eq:BDF2lini2} gives
\begin{equation}\label{eq:BDF2lini3}
\begin{array}{lll}
 \delta_{\omega}^{n+1}-\delta_{\omega}^{n}
   +\frac{1}{2}\xi(\nu\delta_{\omega}^{n+1}+\mu\tilde{\delta}_{ \omega}^{n+1}+\nu\delta_{\omega}^{n}+\mu\tilde{\delta}_{\omega}^{n})=0,
\end{array}
\end{equation}
where both $\mu$ and $\nu$ are constant in time. For SIA it is known that $\frac{\partial \bv}{\partial H_j}\propto \frac{1}{\Delta x}$
and $\mu\propto \frac{1}{\Delta x}$, see \ref{app:b},
and there is reason to believe that $\frac{\partial \bv}{\partial H_j}$ behaves similarly for the FS equation.
Let $\gamma=1/(1+\frac{1}{2}\xi\nu)$ in \eqref{eq:BDF2lini3} to obtain
\begin{equation}\label{eq:BDF2lini4}
\begin{array}{lll}
 \delta_{\omega}^{n+1}+\frac{1}{2}\xi\mu\gamma \tilde{\delta}_{{\omega}}^{n+1}=\gamma(1-\frac{1}{2}\xi\nu)\delta_{\omega}^{n}
   -\frac{1}{2}\xi\gamma\mu\tilde{\delta}_{{\omega}}^{n}.
\end{array}
\end{equation}

AB (\ref{eq:AB}) for $\tilde{H}_{ j}+\tilde{\delta}_{ j}^{n+1}$ is defined by
\begin{equation}\label{eq:AB2lin}
\begin{array}{lll}
 (\tilde{H}_{j}+\tilde{\delta}_{ j}^{n+1})-(H_j+\delta_j^{n})+\\
 \frac{3}{2}\xi\left(\bv_j(\tilde{H}+\tilde{\delta}^{n})(H_j+\delta_j^{n})-\bv_{j-1}(\tilde{H}+\tilde{\delta}^{n})(H_{j-1}+\delta_{j-1}^{n})\right)-\\
     \frac{1}{2}\xi\left(\bv_j(\tilde{H}+\tilde{\delta}^{n-1})(H_j+\delta_j^{n-1})-\bv_{j-1}(\tilde{H}+\tilde{\delta}^{n-1})(H_{j-1}+\delta_{j-1}^{n-1})\right)=a_s.
\end{array}
\end{equation}
After linearization around the steady state, insertion of Fourier modes, and simplification we arrive at
\begin{equation}\label{eq:AB2lin4}
\begin{array}{lll}
 \tilde{\delta}_{\omega}^{n+1}=\delta_{\omega}^{n}
   -\xi(\frac{3}{2}(\nu\delta_{\omega}^{n}+\mu\tilde{\delta}_{\omega}^{n})-\frac{1}{2}(\nu\delta_{\omega}^{n-1}+\mu\tilde{\delta}_{\omega}^{n-1})).
\end{array}
\end{equation}

Stability is investigated by writing the combined scheme AB-FAM with a companion matrix $\fatA$
\begin{equation}\label{eq:stabeq}
  \fatdelta^{n+1}=\fatA\fatdelta^n.
\end{equation}
For stability, the eigenvalues $\lambda_k$ of $\fatA(\mu, \nu, \xi)$ should satisfy $|\lambda_k(\fatA)|\le 1$ for all $k$ and for all $\omega\Delta x\in[0, \pi]$. The companion matrix $\fatA^{2F}(\mu, \nu, \xi)$ and $\fatdelta^n$ for AB-FAM are
\begin{equation}\label{eq:AM2AB2i}
\begin{array}{ll}
\fatA^{2F}=\left(\begin{array}{cccc}
A_{11}&-\frac{1}{2}\xi\gamma\mu(1-\frac{3}{2}\xi\mu)&
                                                   -\frac{1}{4}\xi^2\gamma\mu \nu&-\frac{1}{4}\xi^2\gamma\mu^2\\
1-\frac{3}{2}\xi\nu&-\frac{3}{2}\xi\mu&\frac{1}{2}\xi\nu&\frac{1}{2}\xi\mu\\
1&0&0&0\\
0&1&0&0
\end{array}\right),\\
A_{11}=\gamma(1-\frac{1}{2}\xi\nu-\frac{1}{2}\xi\mu(1-\frac{3}{2}\xi\nu)),\; 
(\fatdelta^n)^T=( \delta_{\omega}^{n},\tilde{\delta}_{\omega}^{n},\delta_{\omega}^{n-1},\tilde{\delta}_{\omega}^{n-1}).
\end{array}
\end{equation}
Here the superscript $2F$ denotes a second order fully implicit scheme.

If we instead consider SAM for $\bH_j+\delta_j^{n+1}$, (\ref{eq:BDF2lini}) is replaced by
\begin{equation}\label{eq:BDF2line}
\begin{array}{l}
 (\bH_j+\delta_j^{n+1})-(\bH_j+\delta_j^{n})+\\
   \frac{1}{2}\xi\left(\bv_j(\tilde{H}+\tilde{\delta}^{n+1})(\tilde{H}_{ j}+\tilde{\delta}_{j}^{n+1})
    -\bv_{j-1}(\tilde{H}+
      \tilde{\delta}^{n+1})(\tilde{H}_{ j-1}+\tilde{\delta}_{ j-1}^{n+1})\right)+\\
   \frac{1}{2}\xi\left(\bv_j(\tilde{H}+\tilde{\delta}^{n})(\bH_j+\delta_j^{n})
  -\bv_{j-1}(\tilde{H}+\tilde{\delta}^{n})(\bH_{j-1}+\delta_{j-1}^{n})\right)=a_s.
\end{array}
\end{equation}

Proceeding as we did for AB-FAM we obtain the following companion matrix $\fatA^{2S}$ for AB-SAM
\begin{equation}\label{eq:AM2AB2e}
\fatA^{2S}=\left(\begin{array}{cccc}
A_{11}&-\frac{1}{2}\xi\mu(1-\frac{3}{2}\xi(\mu+\nu))&
                                                   -\frac{1}{4}\xi^2\nu(\mu+\nu)&-\frac{1}{4}\xi^2\mu(\mu+\nu)\\
1-\frac{3}{2}\xi\nu&-\frac{3}{2}\xi\mu&\frac{1}{2}\xi\nu&\frac{1}{2}\xi\mu\\
1&0&0&0\\
0&1&0&0
\end{array}\right),\\
\end{equation}
with $A_{11}=1-\frac{1}{2}\xi\nu-\frac{1}{2}\xi(\mu+\nu)(1-\frac{3}{2}\xi\nu)$.

The first order methods FE-FBE and FE-SBE have the following companion matrices, respectively,
\begin{equation}\label{eq:EbEf}
\begin{array}{ll}
\fatA^{1F}=\left(\begin{array}{cc}
A_{11}&\xi^2\mu^2\gamma\\
1-\xi\nu&-\xi\mu
\end{array}\right),&
A_{11}=\gamma(1-\xi\mu(1-\xi\nu)),
\end{array}
\end{equation}
\begin{equation}\label{eq:EbEfx}
\begin{array}{ll}
\fatA^{1S}=\left(\begin{array}{cc}
A_{11}&\xi^2\mu(\nu+\mu)\\
1-\xi\nu&-\xi\mu
\end{array}\right),&
A_{11}=1-\xi(\nu+\mu)(1-\xi\nu),
\end{array}
\end{equation}
with $(\fatdelta^n)^T=( \delta_{\omega}^{n},\delta_{\ast\omega}^{n})$.
	
\begin{figure}[H]
\begin{center}
\vspace{-1.3cm}
\includegraphics[width=50mm]{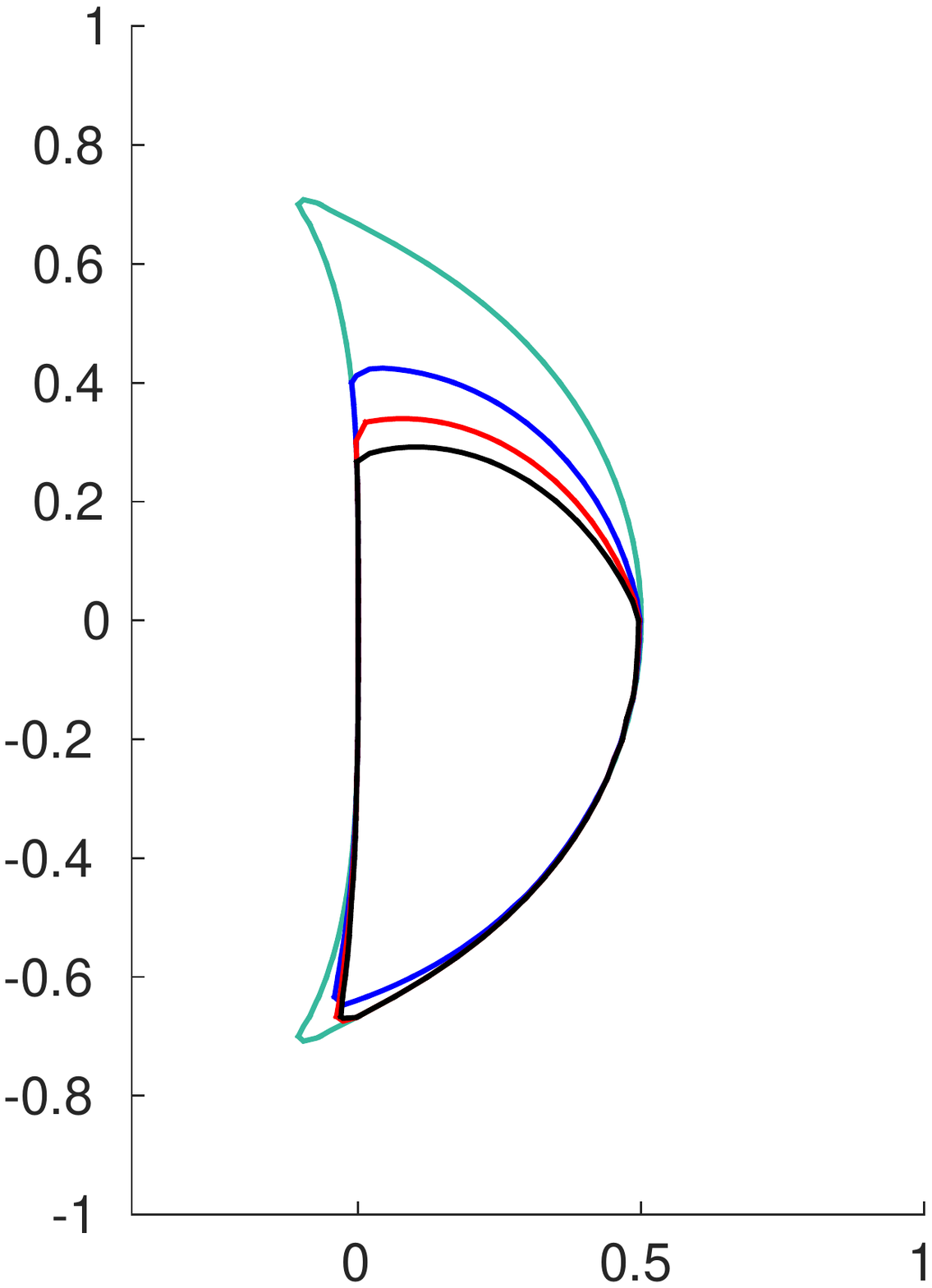} 
\includegraphics[width=50mm]{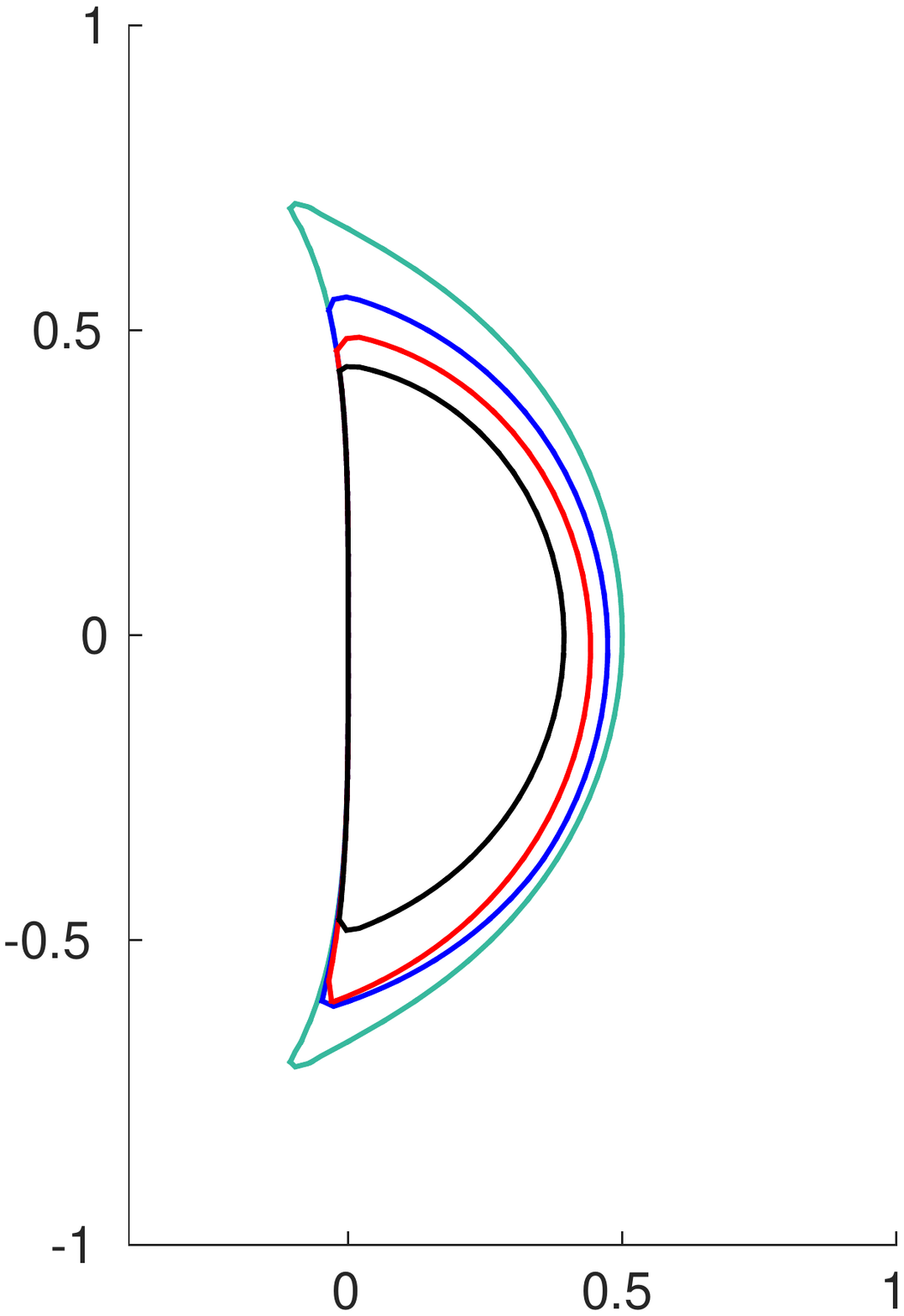} \\

\vspace{-2.5cm}
\includegraphics[width=50mm]{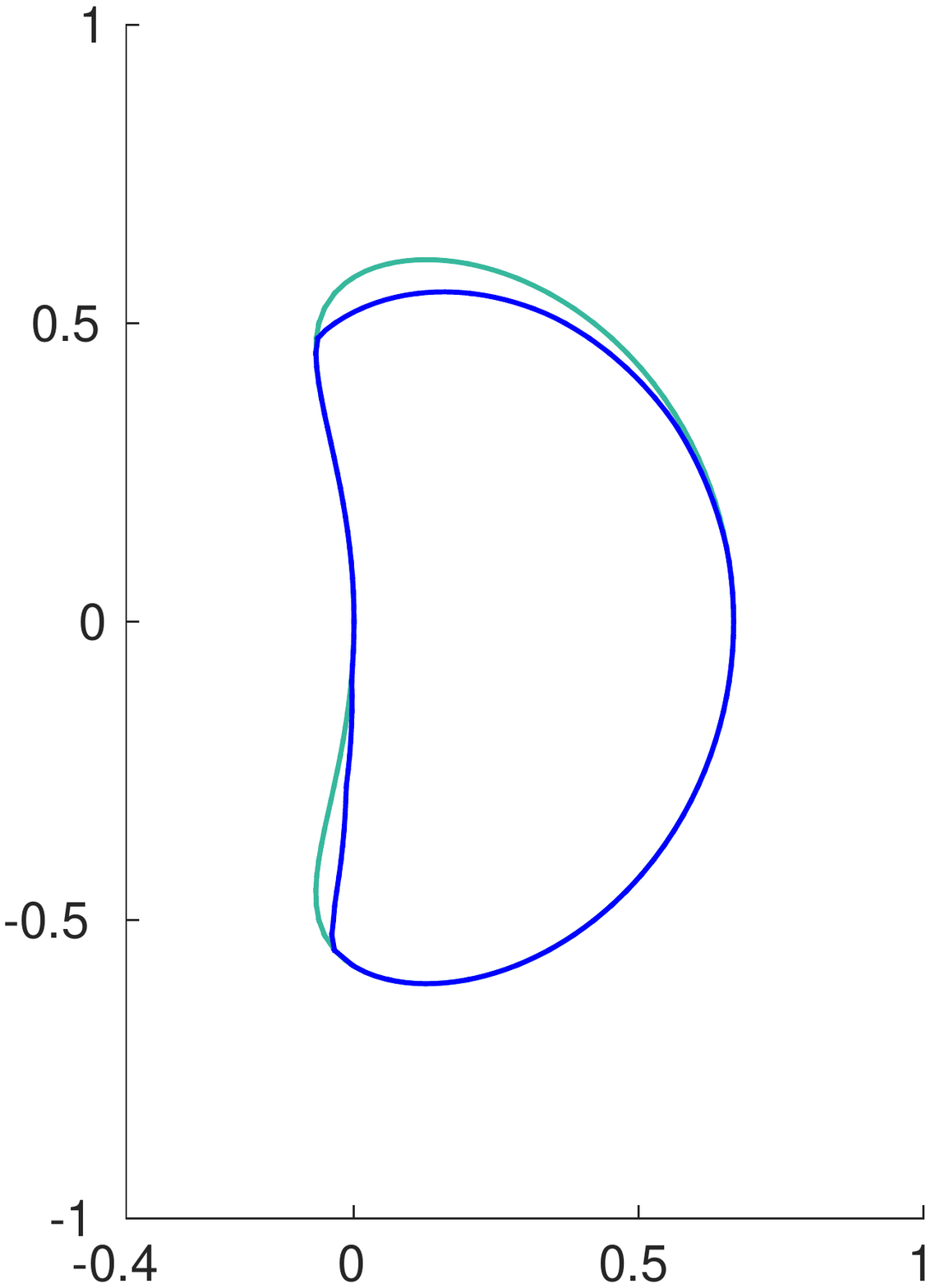} 
\includegraphics[width=50mm]{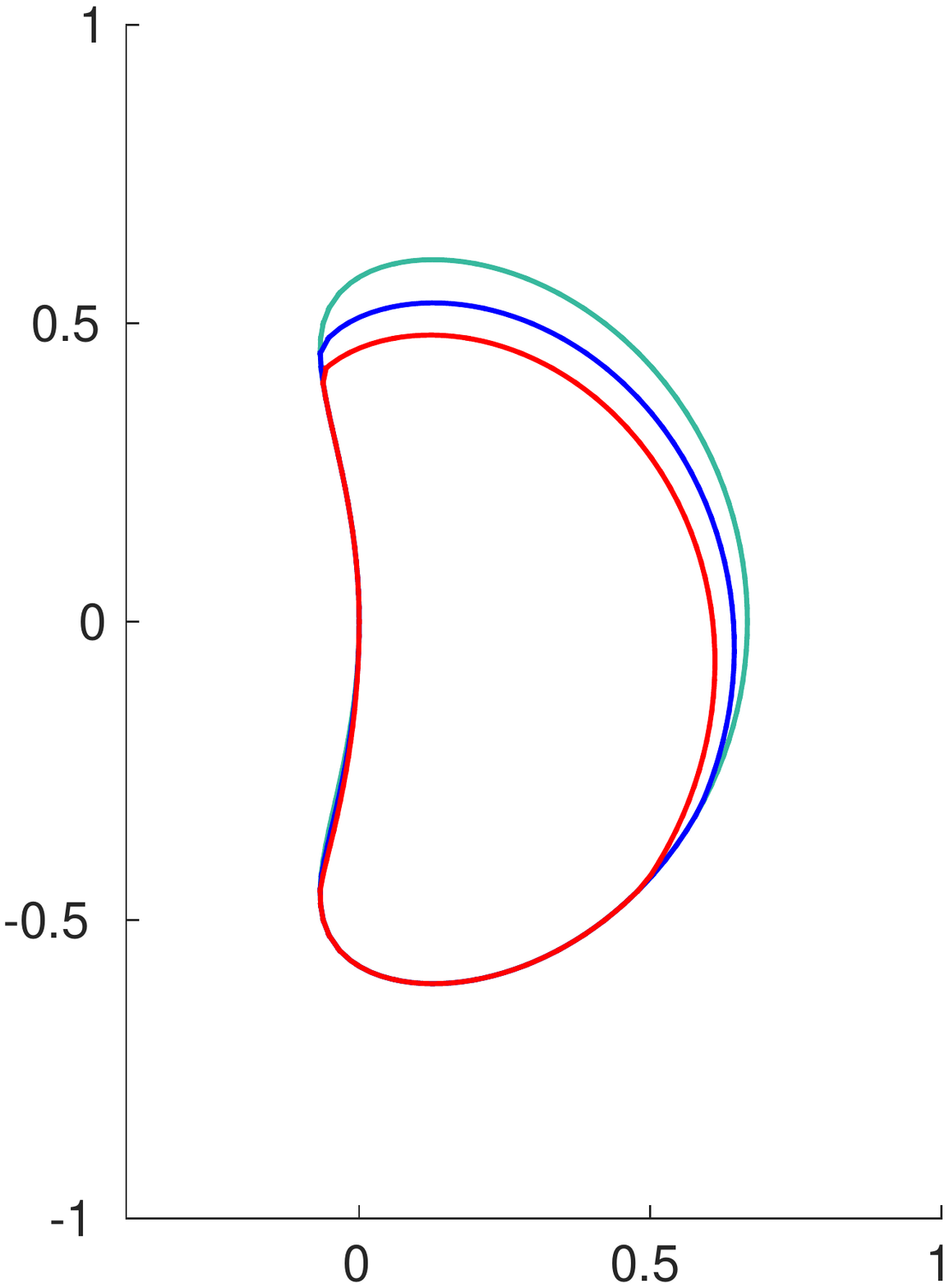} 
\vspace{-1.2cm}
\end{center}
\caption{Stability regions for AB-FAM, AB-SAM, FE-FBE, and FE-SBE in the complex $\xi \mu$-plane for different $\xi \bv$. 
$\Re \xi\mu$ is on the $x$-axis and $\Im \xi\mu$ is on the $y$-axis. The values of $\xi\bv$ increase in the order
green, blue, red, black. Upper row: AB-FAM, $\xi\bv$ is 0,1,2,3, (left),  AB-SAM, $\xi \bv$ is 0,0.2,0.4,0.6, (right). Lower row: FE-FBE, $\xi\bv=0,5$, (left), FE-SBE, $\xi\bv=0,0.2,0.4$, (right).}\label{fig:stabAM2AB2EfEb}

\end{figure} 
The stability regions of AB-FAM, AB-SAM, FE-FBE, and FE-SBE where \\ $\max_{k,\omega\Delta x}|\lambda_k(\fatA(\mu, \nu, \xi)|\le 1$ are found in Figure \ref{fig:stabAM2AB2EfEb}.
The time step $\Delta t$ has to be chosen such that $\xi\mu$ is inside the region for $\xi\bv$.

The stability regions for fully implicit AB-FAM and FE-FBE converge when $\xi\bv$ increases but AB-SAM is 
unstable for all $\mu$ when $\xi\bv> 0.854$ and FE-SBE is unstable for all $\mu$ when  $\xi\bv>0.5$.
Since $\mu\propto 1/\Delta x$ and $\xi\mu\propto 1$ for stability, there is a time step constraint $\Delta t/\Delta x^2\propto 1$ in all methods.  

The predictor-corrector method for time integration is designed such that only one  
solution of the velocity field is necessary in every time step of length $\Delta t_{\rm pc}$. A genuinely implicit method would 
be stable for a longer $\Delta t_{\rm impl}$ but a velocity solution then has to be 
computed in each iteration to solve a system of non-linear
equations involving both $H$ and $\fatv$. Suppose that $n_{\rm it}$ iterations are necessary in the non-linear solver of the thickness 
advection equation and that the work $W_{\rm vel}$ to compute the velocity completely dominates in the
algorithm for ice simulation. The total work in a time interval of length $T$ is then $W_{\rm vel}T/\Delta t_{\rm pc}$ with 
the predictor-corrector scheme and $n_{\rm it}W_{\rm vel}T/\Delta t_{\rm impl}$ 
with the genuinely implicit scheme.

Consider the first order methods FBE in \eqref{eq:FBE} and SBE in \eqref{eq:SBE} with time steps $\Delta t_{\rm impl}$ and $\Delta t_{\rm pc}$, respectively. 
They have the same absolute value of the leading term in the local error $c_{\rm BE}\Delta t^2$ for some $c_{\rm BE}>0$ in \eqref{eq:errSBE}. 
The error tolerance $\epsilon$ is satisfied if $\Delta t_{\rm pc}=\Delta t_{\rm impl}=\sqrt{\epsilon/c_{\rm BE}}$. The implicit FBE has no bound on $\Delta t_{\rm impl}$ for stability.
The stability bound for SBE is $\Delta t_{\rm pc}= c_{\rm stab}\Delta x^2$ for some problem dependent $c_{\rm stab}>0$. Thus, stable and accurate time steps satisfy
\begin{equation}\label{eq:timestepcomp}
   \frac{\Delta t_{\rm pc}}{\Delta t_{\rm impl}}=\min\left(\frac{c_{\rm stab}\Delta x^2\sqrt{c_{\rm BE}}}{\sqrt{\epsilon}}, \; 1 \right).
\end{equation}
If $\Delta x$ is large and $\epsilon$ is small, then $\Delta t_{\rm pc}=\Delta t_{\rm impl}$ and the time steps of both methods are restricted by the accuracy.
On the other hand, when $\Delta t_{\rm pc}/\Delta t_{\rm impl}<1$ then stability bounds $\Delta t _{\rm pc}$. By \eqref{eq:timestepcomp} the 
computational work $W$ for the two methods fulfills
\begin{equation}\label{eq:compwork}
   \frac{W_{\rm impl}}{W_{\rm pc}}=n_{\rm it}\min\left(\frac{c_{\rm stab}\Delta x^2\sqrt{c_{\rm BE}}}{\sqrt{\epsilon}}, \; 1 \right).
\end{equation}
If $\Delta x$ is small and $\epsilon$ large, then the quotient in \eqref{eq:compwork} may be less than 1 and the 
fully implicit FBE is the best choice depending on $n_{\rm it}$ and the problem specific parameters $c_{\rm BE}$ and $c_{\rm stab}$.

\section{Numerical results}
\label{sec:numres}

Three numerical experiments are carried out here with the step size control in Section \ref{sec:adap}. 
The ice is assumed to be a homogeneous isotropic, isothermal material with a constant $\mathcal{A}$ in \eqref{eq:glen}.
The top surface is stress free \eqref{eq:stressfree} and the bottom of the ice is frozen on the bedrock with \eqref{eq:noslip} as the boundary condition. 
The physical parameters of the ice are given in Table \ref{tab: physical param}.
  
\begin{table}[H]
  \begin{center}
    \begin{tabular}{ l l }
      \hline
      Parameter  &  Quantity \\
      \hline
      $\rho=910$~kg m$^{-3}$ & Ice density\\
      $g=9.81$~m s$^{-2}$  & Acceleration of gravity \\
      $\mathcal{A}=100$~MPa$^{-3}$a$^{-1}$ & Rate factor in flow law \\
      \hline
    \end{tabular}
    \caption{The physical parameters of ice.} \label{tab: physical param}
  \end{center}
\end{table}

In Experiment 1, the ice flow is solved by SIA. We compare the four schemes in Table \ref{tab:schemes} and relate the results to the analysis in Section \ref{sec:analysis}. 
In Experiment 2, a 2D flow-line model of an ice sheet is solved by both SIA and FS. The efficiency of the adaptive time stepping method is evaluated in the transient simulation.
Finally in Experiment 3, the adaptive time stepping method is tested on a 3D ice sheet model with FS and ISCAL running for more than 25,000 years.

\subsection{Experiment 1 - 2D slab-on-slope experiment}
\label{sec:numres1}
\subsubsection{Setup}
As in the analysis in Section \ref{sec:analysis}, a 2D slab-on-slope case is considered. Periodical boundary conditions are imposed on the left and right boundaries to represent an infinitely long slab. The computational domain is $L=1000$~km with a uniform mesh size $\Delta x=10$~km. The slope angle is $\alpha=-0.05$ and the steady state thickness of the slab is $H=1000$~m with an initial perturbation $\delta(x)=10\sin(20\pi x/L)$~m at $t=0$. The accumulation rate on the top surface is a constant $a_s=0.3$~m/year or m/a. There is no melting or sliding at the bottom of the slab. 

The velocities are computed by using the analytical solutions of SIA in \eqref{eq:SIA} and the surface evolution is governed by the thickness equation \eqref{eq: Thickness}. We use a finite difference method with an upwind scheme for the spatial discretization since this scheme in this case is identical to the finite element method with the chosen stabilization.

The adaptive time stepping methods in Table \ref{tab:schemes} are implemented in MATLAB in this experiment. The maximal $\Delta t$ is estimated in every time step during the whole simulation according to the analysis in Section \ref{sec: stab analysis}. The maximal $\Delta t$ for the first order methods are estimated numerically by the Fourier analysis of the $\theta$-method in \eqref{eq:lambda4} with $\theta=\frac{1}{5}$, i.e. implicit solution of the thickness equation. The estimated step sizes for the second order methods are computed from a similar analysis  with $\theta=\frac{1}{10}$ in \eqref{eq:lambda4} giving equal weights to the explicit and implicit first derivatives in the thickness equation.

\subsubsection{Results}

\begin{figure}[H]
  \begin{centering}
    \includegraphics[width=0.85\textwidth]{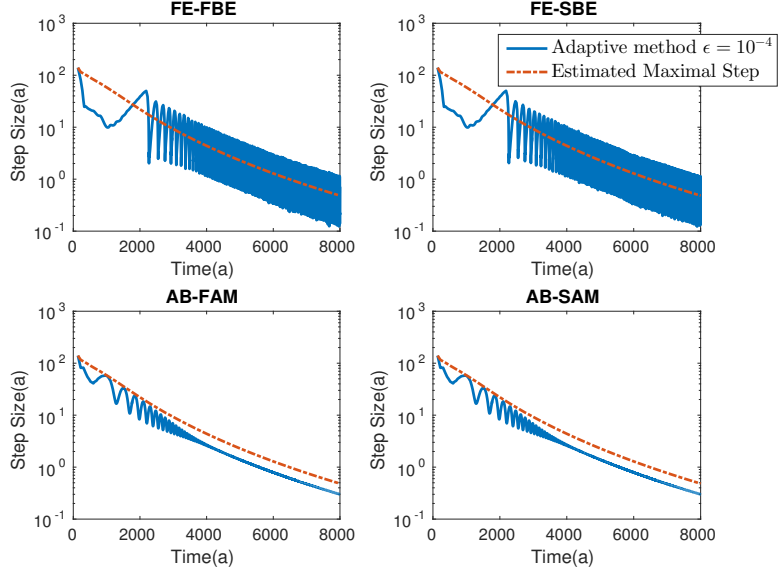}
    \par\end{centering}
  \caption{The time steps of the first and second order methods in the 2D slab-on-slope experiment with a tolerance $\epsilon=10^{-4}$. The actual step sizes taken in the experiments are in solid blue. The estimated maximal step sizes of the corresponding $\theta$-methods are in dashed red with $\theta=1/5$ for FE-FBE and FE-SBE, and  $\theta=1/10$ for AB-FAM and AB-SAM. Upper row: FE-FBE (left) and FE-SBE (right). Lower row: AB-FAM (left) and AB-SAM (right).}\label{fig: adaptive 2d compare}
\end{figure}

The semi-implicit schemes FE-FBE and AB-FAM are compared with the fully implicit schemes FE-SBE and AB-SAM in Figure \ref{fig: adaptive 2d compare}. All four experiments have the same error tolerance $\epsilon=10^{-4}$ and are terminated at $T=8000$ years. The controlled steps follow the estimated maximal steps of the corresponding $\theta$-method. The coupled equation \eqref{eq: h_compact} always contains a diffusion term which is discretized explicitly in time restricting the time step, even for a fully implicit method with $\theta=\frac{1}{5}$ in \eqref{eq:secondorder}.

In this experiment $\xi\bv\approx 10^{-3}$  and $\Re\xi\mu$ oscillates between $10^{-3}$ to $10^1$.
The time steps decrease with increasing time in all cases as the thickness of the ice grows from $1000$~m to about $3400$~m after 8000 years. 
Since $\xi\bv$ is small, the stability bound on $\Delta t$ is such that $\xi\mu\propto 1$ for all methods in Section \ref{sec:stab}.
In \eqref{eq:munu} and \ref{app:b}, $\mu\propto\bH^4$ and consequently
$\Delta t\propto \bH^{-4}$ reducing $\Delta t$ from about $130$ at $t=0$ to $130/3.4^4\approx 1$ at $t=8000$.  
All the four methods in Figure \ref{fig:stabAM2AB2EfEb} are in this case almost at their maximum stability region (shown in green when $\xi\bv$ is small) and 
$\Delta t$ is restricted by $\xi\mu$ in the control method.

There are no significant differences in the stable step sizes between the semi-implicit and fully implicit methods for schemes of the same order. 
However, the semi-implicit methods are computationally cheaper than the fully implicit methods. Therefore, the semi-implicit FE-SBE and AB-SAM methods 
are used in the following experiments for efficiency reasons.

\begin{figure}[H]
  \begin{centering}
    \includegraphics[width=0.85\textwidth]{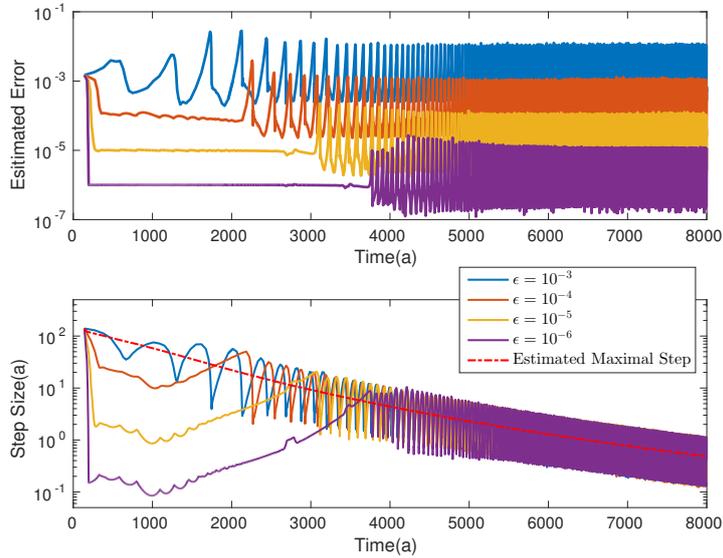}
    \par\end{centering}
  \caption{The 2D slab-on-slope experiment using the first order adaptive time stepping method (FE-SBE). The velocity field is computed by SIA. The maximal step sizes (red dashed) are estimated by using the analysis for $\theta$-method with $\theta=1/5$. The estimated errors (top) are computed by \eqref{eq:tau1}. The step sizes (bottom) are the steps taken in the simulations.}\label{fig: adaptive 2d 1st}
\end{figure}

The time step in the first order methods (FE-SBE and FE-FBE) starts oscillating in magnitude after it reaches the estimated maximal steps. The oscillation is centered around the estimated maximal $\Delta t$. Figure \ref{fig: adaptive 2d 1st} illustrates the estimated errors and step sizes for the FE-SBE method with an error tolerance from $\epsilon=10^{-3}$ to $\epsilon=10^{-6}$. The reference step sizes are computed by the same analysis as in Figure \ref{fig: adaptive 2d compare}. 

For the cases $\epsilon<10^{-3}$, the estimated errors converge to their controlled tolerances immediately by reducing the time step. The estimated error stays constant at $\epsilon$ until $\Delta t$ becomes constrained by the stability condition. The red dashed line indicates the maximal time step for stability by using the $\theta$-method in solving the thickness equation \eqref{eq: H seperated}. During the period where the error stays constant, the stability criterion is always satisfied, and $\Delta t$ is bounded by the accuracy requirement. On the other hand, the time step starts to oscillate when the value of $\xi\mu$ moves out from the stability region of Figure \ref{fig:stabAM2AB2EfEb}. In the case $\epsilon=10^{-3}$, the oscillation starts immediately at the beginning of the simulation since the stability condition is the strongest in the whole interval under this $\epsilon$. 

The reason for the oscillation in the size of the time step is that the adaptive method will try to use longer time steps as long as the estimated accuracy requirements are fulfilled. The PI controller increases $\Delta t$ a few times even if it violates the local stability criterion. The instability will not appear immediately in the controlled error but after a short time the instability is detected in the error estimate and the time step is reduced. 
When the size of the time step starts to oscillate, it is controlled by stability. When $\Delta t$ is bounded by stability it is the same for all values of $\epsilon$ and it follows the shape of the theoretical stability bound.
Note that the size of $\Delta t$ is plotted in a logarithmic scale which means that the absolute amplitude of the oscillation decreases with time.

\begin{figure}[H]
  \begin{centering}
    \includegraphics[width=0.85\textwidth]{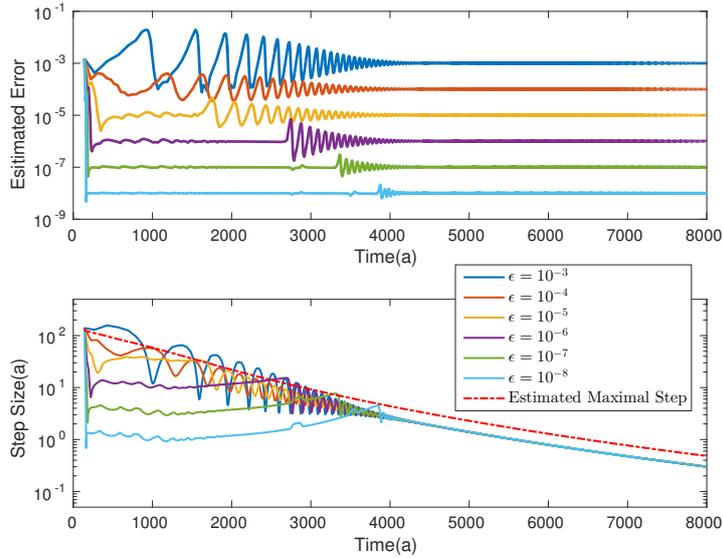}
    \par\end{centering}
  \caption{The 2D slab-on-slope experiment using second order adaptive time stepping method (AB-SAM). The velocity field is computed by SIA. The maximal step sizes (red dashed) are estimated by using the analysis for $\theta$-method. The estimated errors (top) are computed by \eqref{eq:tau2}. The step sizes (bottom) are the steps taken in the simulations.}\label{fig: adaptive 2d 2nd}
\end{figure}

The behavior of the second order method AB-SAM before the oscillations in $\Delta t$ in Figure \ref{fig: adaptive 2d 2nd} is similar to the first order method FE-FBE. The time step is first bounded by the accuracy criterion and then by the stability criterion when the thickness of the ice grows. When the stability bound is reached we see oscillations also here, but the amplitudes are smaller than in the first order case and the oscillations are damped. 

\subsection{Experiment 2 - two dimensional moving margin experiment}

One advantage of using an adaptive time stepping method is the efficiency in transient simulations compared to using a constant minimal $\Delta t$ in the whole interval. Since the time step is automatically adjusted with respect to accuracy and stability, we are able to use as large time steps as possible with respect to both accuracy and stability everywhere in the interval.
\subsubsection{Setup}

In this experiment, a 2D moving margin experiment is performed to test the adaptive time stepping method with FS and SIA in Elmer/Ice. The length of the ice sheet is $L=1000$~km. 
The accumulation rate is defined by
\[
  a_s = \max(0, \min(0.5, s(R-|x-0.5L|))),
\]
where $s=10^{-5}$~a$^{-1}$ and $R=2\times10^5$~m. The initial thickness of the ice is $H(x,0)=100$~m in the whole domain. The mesh size is $\Delta x=1.25$~km in the horizontal direction with $5$ vertical extruded layers which appears to be sufficient in our experiments. FS is solved by the Stokes solver in Elmer/Ice \cite{ElmerDescrip} with a convergence tolerance $10^{-6}$ for the nonlinear system and a direct solver for the linear system. Also, the SIA solver is the one implemented by Ahlkrona in Elmer/Ice \cite{Ahlkrona13}.

\subsubsection{Results}

\begin{figure}[h]
  \begin{centering}
    \includegraphics[width=1\textwidth]{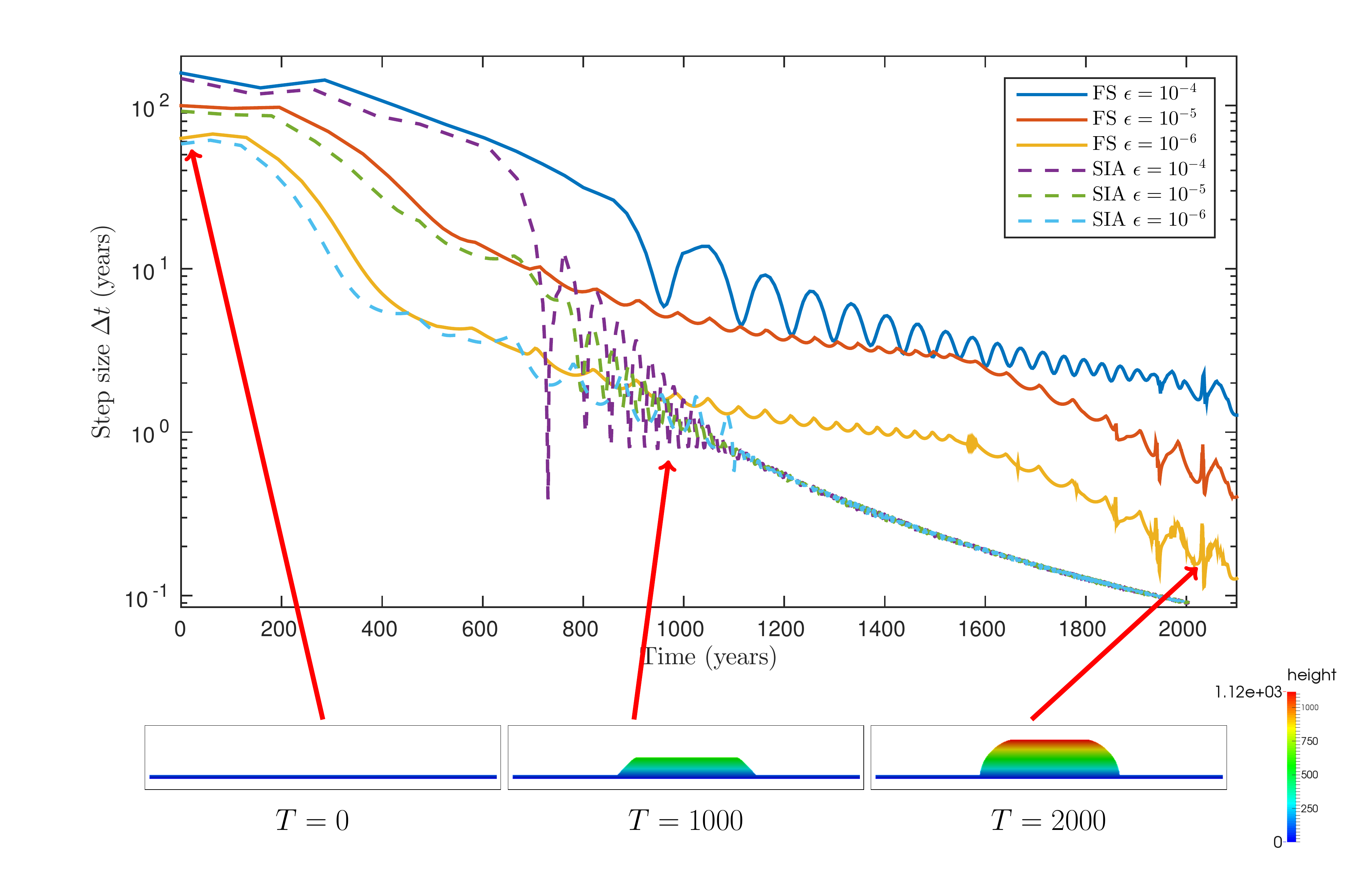}
    \par\end{centering}
  \caption{The 2D moving margin experiment with the AB-SAM method for different tolerances with $\Delta x=1.25$~km solved by Elmer/Ice using SIA or FS. The error tolerance $\epsilon$ used in the time adaptive method is $10^{-4}$, $10^{-5}$, and $10^{-6}$. The three figures at the bottom show the height of the ice cap at three time points. The $y$-axis in these figures is scaled by a factor of $100$ compared to the $x$-axis.} \label{fig: 2D ice cap}
\end{figure}

The second order AB-SAM adaptive time stepping method is run for both SIA and FS system for about $2000$ years in Figure \ref{fig: 2D ice cap}. The three SIA cases behave similarily as in Experiment 1 (Figure \ref{fig: adaptive 2d 2nd}). The size of the time step starts to oscillate as the ice grows thicker, after a while the amplitude of the oscillations decreases, and finally $\Delta t$ converges to the same size for all values of the control $\epsilon$ when the time step is restricted by stability only. The bound on $\Delta t$ in \eqref{eq:dtrestr} decreases when $H$ of the ice cap grows and the slope $\alpha$ of the ice margin increases. In the FS cases, the size of the step is controlled by accuracy in the whole period, while the stability is automatically maintained by the adaptive method. 


The average step sizes of the FS cases are $6.55$, $3.48$ and $1.08$ years for $\epsilon=10^{-4}$, $10^{-5}$, and $10^{-6}$. With a constant time step, $\Delta t$ would have been the smallest one in the interval, i.e. $1.72$, $0.51$ and $0.14$ years, respectively.  The initial time step is reduced by more than two orders of magnitude. Also, we would have had to guess the size of a constant time step to achieve a certain accuracy which is virtually impossible. The efficiency using FS is improved by at least a factor of $4$ (which is the ratio between average step sizes of constant time stepping method and the adaptive method) without losing stability and in control of the accuracy.

\subsection{Experiment 3 - EISMINT 3D with ISCAL}

\subsubsection{Setup}
The capability of handling a 3D ice sheet model is tested in Elmer/Ice in this experiment. The computational domain is $1500\times1500$~km$^2$ with a Cartesian grid ($\Delta x=60$~km) on the horizontal plane and extruded into 5 layers in the vertical direction. The initial thickness of the ice is $H(x,0)=10$~m in the whole domain. The minimal ice thickness is also limited to $10$~m to avoid negative thickness or hanging nodes by melting. The accumulation/ablation rate is 
\[
  a_s = \min(0.5, s(R-d)),
\]
where  
\[
  d = \sqrt{(x-x_\text{center})^2+(y-y_\text{center})^2},
\]
$s=10^{-5}$~a$^{-1}$ and $R=2\times10^5$~m. This is the same configuration as the moving margin experiment in EISMINT 3D benchmark test \cite{Huybrechts,Payne}. 

The problem is first solved by FS with constant time steps ($\Delta t=5$~a) for $12$ initial steps and then by the second order AB-SAM adaptive time stepping method to the steady state. The tolerance is set to $\epsilon=10^{-3}$. Finally, the combination of ISCAL and AB-SAM is tested with the steady state solution of the FS simulation as the spin-up solution. The linear system is solved by the Generalized Conjugate Residual method in Elmer/Ice with a tolerance $10^{-12}$ and the convergence tolerance for non-linear solver is set to $10^{-6}$. The tolerance on the relative error in ISCAL is $5\%$ and the tolerance on the absolute error is $10$~m/a. These tolerances control the switch between the FS and SIA equations in \cite{Ahlkrona16}. 

\subsubsection{Results}

\begin{figure}
  \begin{centering}
    \includegraphics[width=0.9\textwidth]{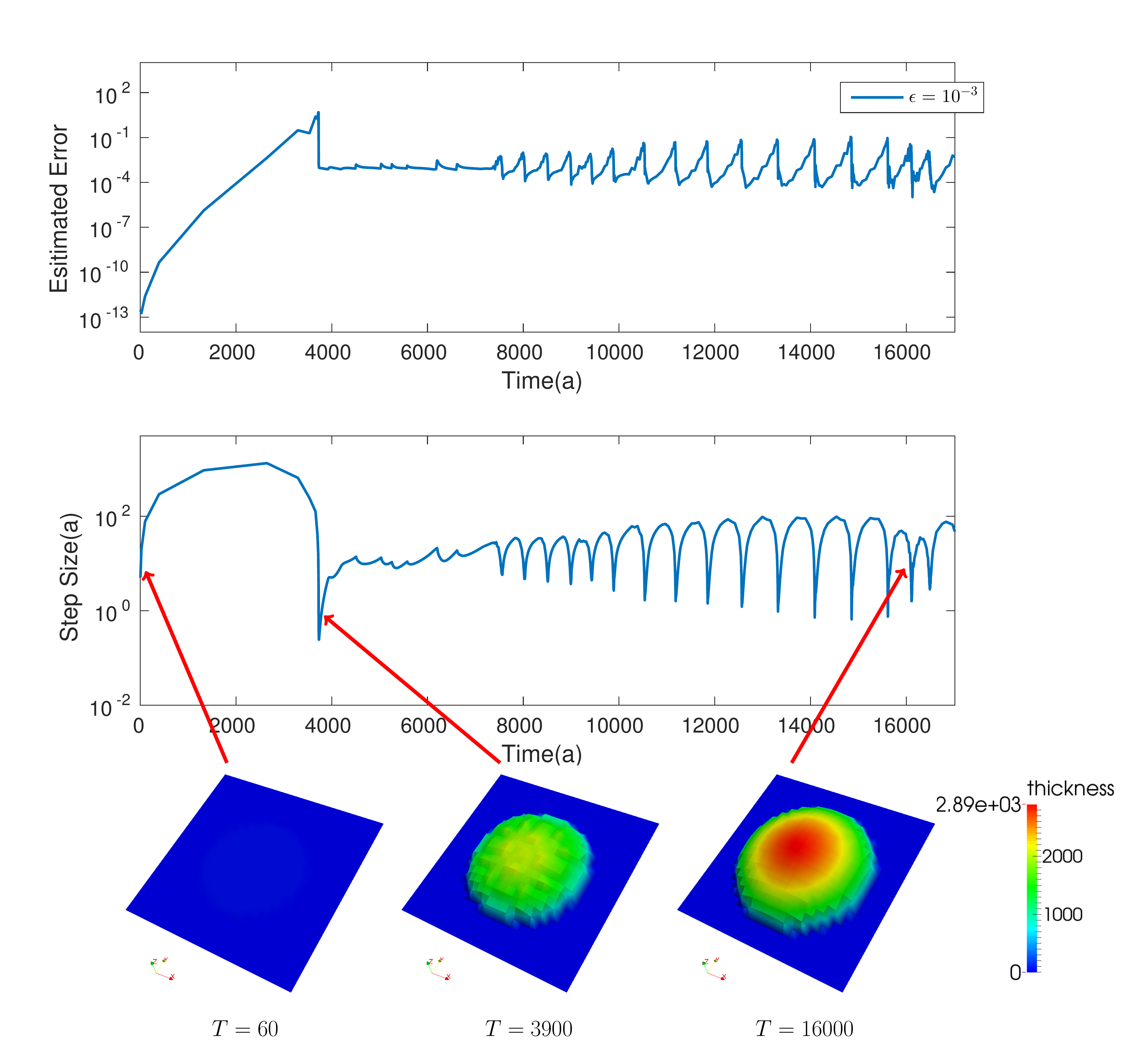}
    \par\end{centering}
  \caption{The EISMINT 3D experiment  ($\Delta x=60$~km)  solved by FS and the second order AB-SAM adaptive time stepping method for the transient simulation. The three figures at the bottom display the thickness of the ice at $T=60$, $3900$ and $16000$ years.}\label{fig: EISMINT 3D FS}
\end{figure}

The result from the transient simulation is shown in Figure \ref{fig: EISMINT 3D FS}. At $T=60$~a, the ice is thin and flat. Therefore, the step size taken by the adaptive method grows quickly and reaches about $1000$~a. At $T\approx3900$~a when the ice cap is formed, the local truncation error exceeds the tolerance. Then, the step size decreases to satisfy stability and accuracy requirements at the steep margin slope (where $\partial h/\partial x$ is large) and the thick ice dome in the ice cap (where $H$ is large) as shown in \eqref{eq:dtrestr}. The estimated local truncation error returns to the tolerance level and oscillates around it until the steady state is reached at $T\approx10000$~a.

\begin{figure}
  \begin{centering}
    \includegraphics[width=0.9\textwidth]{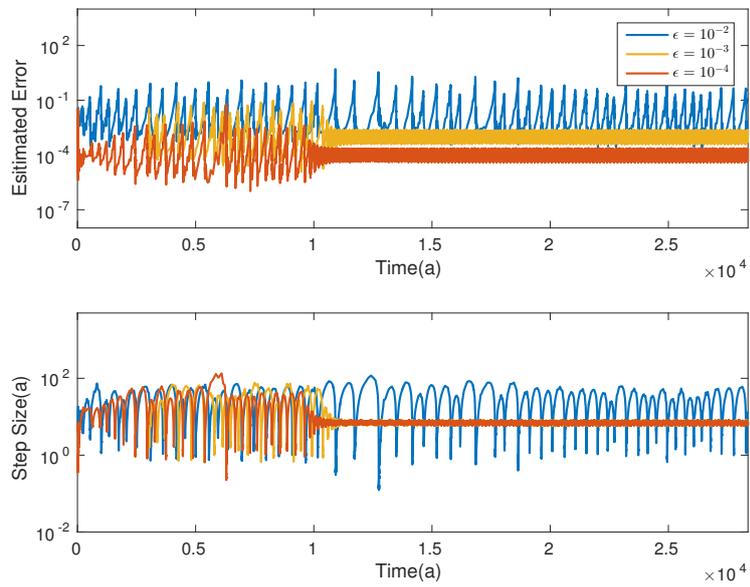}
    \par\end{centering}
  \caption{The EISMINT 3D experiment ($\Delta x=60$~km) with second order AB-SAM adaptive time stepping method for steady states. The problem is simulated by Elmer/Ice using the ISCAL method. The initial condition is the ice cap at $T=16000$ in Figure \ref{fig: EISMINT 3D FS}.}\label{fig: EISMINT 3D}
\end{figure}

The solution from the transient simulation is used as input to a steady-state simulation using ISCAL. The estimated local truncation errors and the time steps are shown in Figure \ref{fig: EISMINT 3D} for the error tolerances $\epsilon=10^{-2}$, $10^{-3}$, and $10^{-4}$. 
The average step size for $\epsilon=10^{-2}$ is $12.41$~years whereas it is $8.10$~years in the case $\epsilon=10^{-3}$ and $7.99$~years for $\epsilon=10^{-4}$. Although the step size oscillates in the whole simulation, the solution is in all cases stable and accurate. 
Combining ISCAL \cite{Ahlkrona16} and our proposed adaptive time stepping method provides a stable, efficient and accurate solution for the steady-state  EISMINT 3D experiment over 25,000 years.

\section{Conclusions and discussion}
\label{sec:conclusions}
With a constant time step $\Delta t$ in the whole time interval of interest, 
we have to take the minimal one in the interval for stability. The accuracy of the approximation of the time derivative
is difficult to assess {\it a priori} with a fixed time step. Instead, a time step control is proposed and tested here for the  
shallow ice approximation (SIA), the full Stokes (FS) equation, the combination of them in ISCAL, and the thickness advection
equation. The stability of the numerical solution is maintained and the accuracy is controlled by keeping the local error below a given threshold. Depending on the threshold,
$\Delta t$ is bound by stability requirements or accuracy requirements. 

The most expensive part of the simulations is to determine the velocity field in the ice in each time step. 
To solve the FS equations is
very costly, ISCAL is less expensive, and SIA is fairly cheap but still much more computationally expensive than solving the thickness advection equation in one dimension lower.
We have developed a method advancing the solution in time and estimating the time discretization error requiring only one solution of the velocity field per time step. 
The method takes a shorter $\Delta t$ than an implicit method but with less work in each time step and the solver is simpler.
Our method is applied to the simulation of a 2D ice slab and a 3D circular ice sheet.
The stability bounds in the experiments are explained by and agree well with the theoretical results.

\section*{Acknowledgment}
This work has been supported by FORMAS grant 2013-1600 to Nina Kirchner.
The computations were performed on resources provided by SNIC through
Uppsala Multidisciplinary Center for Advanced Computational Science
(UPPMAX) and PDC Center for High Performance Computing at KTH. Thanks to Josefin Ahlkrona for careful reading of a draft of the paper.

\bibliographystyle{elsarticle-num}
\bibliography{stab}

\appendix

\section{Stability criterion for simplified 2D-problem}
\label{app:a}
Let $\kappa=\omega\Delta x$ and $a$ replace $|a|$ in \eqref{eq: growth factor 2} and compute $|\lambda|^2$ for $\lambda$ 
\begin{equation} 
  \begin{array}{rcl}
    |\lambda|^2&=&(a-2c)^2+(1-a-2c)^2+8c^2\cos^3{\kappa}+4c(1-2c)\cos^2{\kappa}+\\
    &+&(2a-2a^2-8c^2)\cos{\kappa}\\
    &=&8c^2\cos^3{\kappa}+(4c-8c^2)\cos^2{\kappa}+(2a-2a^2-8c^2)\cos{\kappa}+\\
    &+&2a^2+8c^2-2a-4c+1.\\
  \end{array}
\end{equation}
To obtain $|\lambda|^2\leq 1$ we need
\begin{equation} 
    (\cos{\kappa}-1)(8c^2\cos^2{\kappa}+4c\cos{\kappa}+2a-2a^2-8c^2+4c)\leq 0,
\end{equation}
i.e. either $\cos{\kappa}=1$ or 
\begin{equation} 
  8c^2\cos^2{\kappa}+4c\cos{\kappa}+2a-2a^2-8c^2+4c\geq 0.
\end{equation}
By denoting
\begin{equation} 
  g(\kappa)=8c^2\cos^2{\kappa}+4c\cos{\kappa}+2a-2a^2-8c^2+4c\label{eq: simplified obj 1},
\end{equation}
the problem becomes to find $\kappa^*$ that minimizes $g(\kappa)$ and then solve $|\lambda|^2\leq1$. Since $a>0,\,c>0$, the minimizing problem has the solution
\[
  \cos{\kappa^*}=-\frac{1}{4c},
\]
and the inequality to solve for $\Delta t$ becomes
\begin{equation} 
  8c^2-4c+2a^2-2a+\frac{1}{2}\leq 0.
  \label{eq:ac}
\end{equation}
We only need to consider the case  where $c\geq\frac{1}{4}$, since for $c<\frac{1}{4}$ the inequality holds for all $\Delta t$. Let $a = kc$ for $k>0$ which in (\ref{eq:ac}) gives
\begin{equation} 
  (8+2k^2)c^2-(4+2k)c+\frac{1}{2}\leq 0,
\end{equation}
i.e.
\begin{equation} 
  \frac{k+2-2\sqrt{k}}{2k^2+8}\leq c\leq\frac{k+2+2\sqrt{k}}{2k^2+8}.
\end{equation}
Since $c\geq\frac{1}{4}$ and $k>0$ only the right inequality becomes a restriction and we get 
\begin{equation}  
  c\leq\frac{k+2+2\sqrt{k}}{2k^2+8},
\end{equation}
which gives
\begin{equation}  
  \Delta t\leq\frac{k+2+2\sqrt{k}}{2k^2+8}\frac{2\Delta x^2}{3C\alpha^2\bH^5}.
\end{equation}

\section{Dependence of $\bv$ on $H$ in the discretization}
\label{app:b}

The formula for $\bv$ at $x_j$ for SIA in \eqref{eq: average velocity} with a discretized first derivative can be written
\begin{equation}  \label{eq:vSIA2}
   \bv= -C H^4(\calD h)_j^3 =(\calD h)_j^3 G(H),
\end{equation}
where $(\calD h)_j$ is the discrete approximation of $\partial h/\partial x$ at $x_j$. Then the sensitivity
$\partial \bv/\partial H_j$ in Section \ref{sec:stab} is
\begin{equation}  \label{eq:sensvSIA}
   \displaystyle{\frac{\partial \bv}{\partial H_j}=
    (\calD h)_j^3 \frac{G(H_j)}{\partial H_j}+\left(3(\calD h)_j^2\frac{\partial (\calD h)_j}{\partial h_j}\frac{\partial h_j}{\partial H_j}\right) G(H_j)}.
\end{equation}
All components on the right hand side of \eqref{eq:sensvSIA} are smooth and of $\ordo 1$ except for ${\partial (\calD h)_j}/{\partial h_j}$ which is of $\ordo{\Delta x^{-1}}$.
Therefore, $\partial \bv/\partial H_j$ and $\mu$ in \eqref{eq:munu} are both of $\ordo{\Delta x^{-1}}$ and potentially large for small $\Delta x$.

\end{document}